\documentclass{sig-alternate}
\usepackage{latexsym}
\usepackage{times}
\usepackage{epsfig}
\usepackage{subfigure}
\usepackage{multirow}
\usepackage{amsfonts}
\usepackage{algorithm,algorithmic}
\usepackage[usenames]{color}
\usepackage{graphicx}
\usepackage{caption}
\usepackage{url}
\usepackage{exscale,relsize}

\newcommand{\x}[1]{{\color{red}\bf{}}}

\newcommand{\comment}[1]{}

\newtheorem{example}{Example}[section]

\newsavebox{\boxone}
\newsavebox{\boxtwo}
\newsavebox{\boxthree}

\newlength{\narrow}
\newlength{\cnarrow}
\newcommand{\topline}{
  \hrule
  \vskip .5\baselineskip}
\newcommand{\bottomline}{
  \vskip 2pt
  \hrule}

\newcommand{\chbox}[2]{
  \hbox to #1{\hss\vtop{#2}\hss}}

\newcommand{\nchbox}[1]{
  \chbox{\narrow}{#1}}

\newcommand{\cnchbox}[1]{
  \chbox{\cnarrow}{#1}}

\newcommand{\fcode}[1]{
  
  \chbox{\textwidth}{\tgrind\input{#1}}}

\newcommand{\ncode}[1]{
  
  \chbox{\narrow}{\tgrind\input{#1}}}

\def\nfig#1#2#3{
  \vtop{\nchbox{#1}
  \hbox to\narrow{\parbox{\narrow}{\caption{#2}\label{#3}}}}}

\newcommand{\cncode}[1]{
  \chbox{\cnarrow}{\tgrind\input{#1}}}

\def\codefiggen[#1]#2#3#4#5#6{
  \begin{figure}[#1]
  #5
  \fcode{#2}
  \center\parbox{.9\textwidth}{\caption{#3}\label{#4}}
  #6
  \end{figure}}

\def\codefig[#1]#2#3#4{
  \codefiggen[#1]{#2}{#3}{#4}{}{}}

\def\codefigline[#1]#2#3#4{
  \codefiggen[#1]{#2}{#3}{#4}{\topline}{\bottomline}}

\def\doublefiggen[#1]#2#3#4#5#6#7#8#9{
  \begin{figure}[#1]
  #8
  \hbox to \textwidth{
  \nfig{#2}{#3}{#4}
  \hfil
  \nfig{#5}{#6}{#7}}
  #9
  \end{figure}}

\def\doublefig[#1]#2#3#4#5#6#7{
  \doublefiggen[#1]{#2}{#3}{#4}{#5}{#6}{#7}{}{}}

\def\doublefigline[#1]#2#3#4#5#6#7{
  \doublefiggen[#1]{#2}{#3}{#4}{#5}{#6}{#7}{\topline}{\bottomline}}

\def\doublecodefig[#1]#2#3#4#5#6#7{
  \doublefig[#1]{\ncode{#2}}{#3}{#4}{\ncode{#5}}{#6}{#7}}

\def\doublecodefigline[#1]#2#3#4#5#6#7{
  \doublefigline[#1]{\ncode{#2}}{#3}{#4}{\ncode{#5}}{#6}{#7}}

\newcommand{\codepair}[4]{\vbox{
  \hbox{\ncode{#1} \hfil \ncode{#3}}
  \vskip .3\baselineskip plus .3\baselineskip
  \hbox{\hbox to\narrow{#2\hfil} \hfil \hbox to\narrow{#4\hfil}}}}

\def\codepairfig[#1]#2#3#4#5#6#7{
  \begin{figure}[#1]
  \codepair{#2}{#3}{#4}{#5}
  \center\parbox{.9\textwidth}{\caption{#6}}
  \label{#7}
  \end{figure}}

\def\cncodepairfiggen[#1]#2#3#4#5#6#7{
  \begin{figure}[#1]
  #6
  \hbox{\cncode{#2}\hfil\cncode{#3}}
  \center\parbox{.9\columnwidth}{\caption{#4}\label{#5}}
  #7
  \end{figure}}

\def\cncodepairfig[#1]#2#3#4#5{
  \cncodepairfiggen[#1]{#2}{#3}{#4}{#5}{}{}}

\def\cncodepairfigline[#1]#2#3#4#5{
  \cncodepairfiggen[#1]{#2}{#3}{#4}{#5}{\topline}{\bottomline}}

\def\doublefigOnecap*[#1]#2#3#4#5{
  \begin{figure*}[#1]
  \hbox to \textwidth{
  \nchbox{#2}
  \hfil
  \nchbox{#3}}
  \caption{#4}
  \label{#5}
  \end{figure*}}

\def\doublefigOnecap[#1]#2#3#4#5{
  \begin{figure}[#1]
  \topline
  \hbox to \columnwidth{
  \cnchbox{#2}
  \hfil
  \cnchbox{#3}}
  \caption{#4}
  \label{#5}
  \bottomline
  \end{figure}}

\def\PSfig[#1]#2#3#4{
 \begin{figure}
 \centerline{\psfig{file=#2,width=\columnwidth}}
 \caption{{#3}}
 \label{#4}
 \end{figure}}

\def\PSfiglines[#1]#2#3#4{
 \begin{figure}[#1]
 \topline
 \centerline{\psfig{file=#2,width=\columnwidth}}
 \caption{{#3}}
 \label{#4}
 \bottomline
 \end{figure}}

\def\PSfiglinesht[#1]#2#3#4#5{
 \begin{figure}[#1]
 \topline
 \centerline{\psfig{file=#2,height=#3}}
 \caption{{#4}}
 \label{#5}
 \bottomline
 \end{figure}}

\def\doublePSfig[#1]#2#3#4#5#6{
  \doublefigOnecap[#1]
    {\cnchbox{\psfig{file=#2,height=#4}}}
    {\cnchbox{\psfig{file=#3,height=#4}}}
    {#5}
    {#6}}

\newlength{\boxwidth}
\newcommand{\bproof}{{\bf Proof Sketch: }}
\newcommand{\eproof}{\mbox{$\Box$}}

\def\tabdoublecode#1#2#3#4{
 \begin{figure*}[t]
 \topline\vs{-.4}
 \hbox to \columnwidth{
 \vtop{\small
 \begin{tabbing}
 #1
 \end{tabbing}}
 \hfil
 \hfil
 \hfil
 \vtop{\small
 \begin{tabbing}
 #2
 \end{tabbing}}
 }
 \caption{#3\label{#4}}
 \bottomline
 \end{figure*}
}
\def\tabtriplecode#1#2#3#4#5{
 \begin{figure}
 \topline\vs{-.4}
 \hbox to \columnwidth{
 \vtop{\small
 \begin{tabbing}
 #1
 \end{tabbing}}
 \hfil
 \hfil
 \hfil
 \vtop{\small
 \begin{tabbing}
 #2
 \end{tabbing}}
 \hfil
 \hfil
 \hfil
 \vtop{\small
 \begin{tabbing}
 #3
 \end{tabbing}}
 }
 \caption{#4\label{#5}}
 \bottomline
 \end{figure}
}

\newtheorem{lemma}{Lemma}
\newcommand{\blemma}{\begin{lemma}}
\newcommand{\elemma}{\end{lemma}}

\newtheorem{thm}{Theorem}
\newcommand{\bthm}{\begin{thm}}
\newcommand{\ethm}{\end{thm}}

\newtheorem{defin}{Definition}
\newcommand{\bdefin}{\begin{defin}}
\newcommand{\edefin}{\end{defin}}

\newtheorem{observation}{Observation}
\newcommand{\bobserv}{\begin{observation}}
\newcommand{\eobserv}{\end{observation}}

\newcommand{\vs}[1]{\vspace{#1cm}}
\newcommand{\be}{\begin{equation}}
\newcommand{\ee}{\end{equation}}

\newcommand{\bdesc}{\begin{description}}
\newcommand{\edesc}{\end{description}}
\newcommand{\benum}{\begin{enumerate}}
\newcommand{\eenum}{\end{enumerate}}
\newcommand{\bitem}{\begin{itemize}}
\newcommand{\eitem}{\end{itemize}}
\newcommand{\bcenter}{\begin{center}}
\newcommand{\ecenter}{\end{center}}
\newcommand{\btabular}{\begin{tabular}}
\newcommand{\etabular}{\end{tabular}}
\newcommand{\beqnarr}{
 \begin{eqnarray}}
\newcommand{\eeqnarr}{\end{eqnarray}}

\makeatletter
\newcommand{\raisemath}[1]{\mathpalette{\raisem@th{#1}}}
\newcommand{\raisem@th}[3]{\raisebox{#1}{$#2#3$}}
\makeatother

\title{Hub-Accelerator: Fast and Exact Shortest Path Computation in Large Social Networks}

\numberofauthors{1}
\author{\alignauthor Ruoming Jin$^\dagger$ ~~~~~ Ning Ruan$^\star$ ~~~~~~ Bo You$^\dagger$ ~~~~~~ Haixun Wang$^\ddagger$ \\
\affaddr{$^\dagger$\mbox{ }Kent State University ~~~~~~~~~~~~ $^\star$ \mbox{ } Google Inc ~~~~~~~~~ $^\ddagger$ \mbox{ } Microsoft Research Asia } \\
\email{\{jin,byou\}@cs.kent.edu ~~~~~~~~ ningruan@google.com ~~~~~~haixunw@microsoft.com}
}

\begin{document}



\maketitle

\begin{abstract}

Shortest path computation is one of the most fundamental operations for managing and analyzing large social networks.
Though existing techniques are quite effective for finding the shortest path on large but sparse road networks,
social graphs have quite different characteristics: they are generally non-spatial, non-weighted, scale-free, and they exhibit small-world properties in addition to their massive size. 
In particular, the existence of hubs, those vertices with a large number of connections, explodes the search space, making the shortest path computation surprisingly challenging.
In this paper, we introduce a set of novel techniques centered around hubs, collectively referred to as the Hub-Accelerator framework, to compute the $k$-degree shortest path (finding the shortest path between two vertices if their distance is within $k$). These techniques enable us to significantly reduce the search space by either greatly limiting the expansion scope of hubs (using the novel {\em distance-preserving Hub-Network} concept) or completely pruning away the hubs in the online search (using the {\em Hub$^2$-Labeling} approach). 
The Hub-Accelerator approaches are more than two orders of magnitude faster than BFS and the state-of-the-art approximate shortest path method Sketch for the shortest path computation. 
The Hub-Network approach does not introduce additional index cost with light pre-computation cost;  the index size and index construction cost  of Hub$^2$-Labeling are also moderate and better than or comparable to the approximation indexing  Sketch method. 

\comment{
Computing the shortest path between any two vertices plays a fundamental role on many applications, ranging from semantic web, ontology processing and social network analysis.
Recently, it faces a great challenge arisen by the emergence of massive networks, e.g., social networks.
Most of existing techniques perform well on medium-sized or relatively sparse graph, such as road network, become infeasible to them.
Moreover, though some recent methods attempt to efficiently approximate the distance or shortest paths, finding {\em exact} shortest paths is still known to be unattainable.
In this work, we present a new approach, called SPARSE, which can efficiently answer {\em exact} shortest path queries in a large social network. 
Its query efficiency is supported by a novel landmark-based indexing scheme, which ont only produces compact indices, but also can help significantly improve online search.
Interestingly, by varying the number of landmarks, we can easily trade off the query performance and memory assumption for indexing.
The comprehensive experimental evaluation demonstrates that our SPARSE can effectively handle graphs with up to 10 millions vertices and
its query processing is much faster than the state-of-the-art shortest path approximation methods.
}

\comment{
Most of existing techniques which work well on medium-size or relatively sparse graphs, such as road network, face performance bottleneck on newly emerging massive network, e.g., social networks.

Most of the existing reachability indices perform well on small- to medium- size graphs, but reach a scalability bottleneck around one million vertices/edges.
As graphs become increasingly large, scalability is quickly becoming the major research challenge for the reachability computation today.
Can we construct indices which scale to graphs with tens of millions of vertices and edges?
Can the existing reachability indices which perform well on moderate-size graphs be scaled to very large graphs?
In this paper, we propose {\bf SCARAB} (standing for SCAlable ReachABility), a unified reachability computation framework:
it not only can scale the existing state-of-the-art reachability indices, which otherwise could only be constructed and work on moderate size graphs, but also can help speed up the online query answering approaches.
Our experimental results demonstrate that SCARAB can perform on graphs with millions of vertices/edges and is also much faster then GRAIL, the state-of-the-art scalability index approach.
}
\end{abstract}

\vspace*{-2.0ex}
\section{Introduction}
\label{intro}

Social networks are becoming ubiquitous and their data volume is increasing dramatically.
The popular online social network websites, such as Facebook, Twitter, and LinkedIn, all have hundreds of millions of active users nowadays.
Google's new social network Google+ attracted 25 million unique users and was growing at a rate of roughly one million visitors per day in the first month after launch.
Enabling online and interactive query processing of
these massive graphs, especially to quickly capture and discover the relationship between entities,  is becoming an indispensable component for emerging applications ranging from the social sciences to advertisement and marketing research, to homeland security.

Shortest path computation is one of the most basic yet critical problems for managing and querying social networks.
The social network website LinkedIn  pioneered the well-known shortest-path service ``How you're connected to A'', which offers a precise description of the friendship chain between you and  a user A within $3$ steps.
Microsoft's Renlifang (EntityCube) ~\cite{entitycube}, which records over a billion relationships for over
10 million entities (people, locations, organizations), allows users to retrieve the shortest path between two entities if their distance is less than or equal to $6$.
The newly emerged online application ``Six Degrees''~\cite{6degreeonline} provides an interactive way to demonstrate how you connect to other people in your Facebook network.
In addition, shortest path computation is also useful in determining trust and discovering friends in online games~\cite{citeulike:7372248,zhao2011}.

In this paper, we investigate the {\bf $k$-degree shortest path} query ($k \leq 6$ in general), which can be formally described as: 
{\em Given two vertices (users) $s$ and $t$ in a large (social) network, what is the shortest path from $s$ to $t$ if their distance is less than or equal to $k$?}
In all these emerging social network applications, (one) shortest path between two users needs to be computed generally only if their distance is less than a certain threshold (such as $6$).
Such a focus directly resonates with the {\em small-world} phenomenon being observed in these massive social networks.
For instance, the average pairwise distance on a large sample of Facebook users~\cite{6degreeonline} has been shown to be only $5.73$. Also, around half the users on Twitter are on average $4$ steps away from another while nearly everyone is $5$ steps away ~\cite{sysomos}.
Not only are most of the users in large social networks separated by less than $6$ steps, the longer connections or paths in social networks are also less meaningful and/or useful.

Computing $k$-degree shortest path in a large social network is surprisingly challenging, especially when $k$ is relatively large, such as $k=6$.
A single BFS (Breadth-First-Search) can easily visit more than a million vertices in $6$ steps in a large network with a few million of vertices.
Though existing techniques~\cite{Jing98,Jung02,Shekhar97,Sanders05,BastFSS07,Gutman04,GoldbergSODA05,HananSA08,Sankaranarayanan:2009,Kriegel:2008:HGE,Geisberger:2008:CHF,Tao11,Bauer:2010:CHG} are very effective for finding the shortest path on large but sparse road networks,
social graphs have quite different characteristics.
Instead of being spatial, with edge weight, and having low vertex degree, social networks are generally {\em non-spatial}, {\em non-weighted}, {\em scale-free} (therefore containing high-degree hub nodes), and they exhibit {\em small-world} properties  in addition to their massive size.
Indeed, due to the difficulty in finding the shortest path in social networks,  the recent studies~\cite{Gubichev:2010:FAE:1871437.1871503,citeulike:7372248,zhao2011} all focus on discovering only the approximate ones (longer than the true shortest path).
Furthermore, even with the approximation, the fastest methods, such as {\em Sketch}~\cite{Gubichev:2010:FAE:1871437.1871503}, {\em TreeSketch}~\cite{Gubichev:2010:FAE:1871437.1871503}, and {\em RigelPaths}~\cite{zhao2011}, still need tens or hundreds of milliseconds ($10^{-3}$ second) to compute an approximate shortest path in a social network with a few million vertices.

The central problem of shortest path computation in massive social network comes from {\em hubs}: those vertices with a large number of connections. The number of hubs may be small compared to the total network size; however, they appear in the close neighborhood of almost any vertex. Indeed, hubs play a critical role in the small-world (social) networks; they serve as the common mediators linking the shortest path between vertices, just like the hub cities in the small-world network of airline flight. In fact, theoretical analysis shows that a small number of hubs (due to the power law degree distribution) significantly shortens the distance between vertices and makes networks ``ultra-small''~\cite{CohenHavlin03}. However, hubs are the key contributing factor to the search-space explosion. Assuming a hub has $5,000$ friends and normal persons have about $100$ friends, then a two-step BFS from the hub will visit $\approx 500,000$ vertices; in the Twitter network, some vertices (celebrities) contain more than $10$ million followers, so a reverse one-step BFS (from that vertex to its followers) is already too expensive. Thus, hubs are at the center of the problem: shortest paths do not exist without them; but they make the discovery extremely hard. Can we disentangle the love-hate relationship between shortest path and hubs? Can we make hubs more amicable for shortest path computation?

In this paper, we provide a positive answer to these challenging problems on shortest path computation in massive social graphs. We introduce a list of novel techniques centered around hubs, collectively referred to as the Hub-Accelerator framework. These techniques enable us to significantly reduce the search space by either greatly limiting the expansion scope of hubs (using the novel {\em distance-preserving hub-network} concept) or completely pruning away the hubs in the online search (using the {\em Hub$^2$-labeling} approach). 
The Hub-Accelerator approaches are on average  more than two orders of magnitude  faster than the BFS and the state-of the-art approximate shortest path methods, including {\em Sketch}~\cite{Gubichev:2010:FAE:1871437.1871503}, {\em TreeSketch}~\cite{Gubichev:2010:FAE:1871437.1871503}, and {\em RigelPaths}~\cite{zhao2011}.
The Hub-Network approach does not introduce additional index cost with light pre-computation cost;  the index size and index construction cost  of Hub$^2$-Labeling are also moderate and better than or comparable to the approximation indexing  Sketch method. 
We note that though the shortest path computation has been extensively studied, most of the studies only focus on road networks ~\cite{Jing98,Jung02,Shekhar97,Sanders05,BastFSS07,Gutman04,GoldbergSODA05,HananSA08,Sankaranarayanan:2009,Kriegel:2008:HGE,Geisberger:2008:CHF,Tao11,Bauer:2010:CHG,Abraham:2010:HDS,Abraham:2011:HLA} or approximate shortest path (distance) computation on massive social networks~\cite{Gubichev:2010:FAE:1871437.1871503,zhao2011}.  To our best knowledge, this is the first work explicitly addressing the exact shortest path computation in these networks.   The Hub-Accelerator techniques are also novel and the distance-preserving subgraph (hub-network) discovery problem itself is of both theoretical and practical importance for graph mining and management.


\comment{
The rest of papers are organized as follows.
We first review the related works on shortest path computation (Section~\ref{related}).
Then, we introduce the  Hub$^2$-Labeling framework (Section~\ref{framework}) and the indexing method (Section~\ref{hubsetselectionandlabeling}).
We discuss the query algorithm in Section~\ref{query}.
We present the detailed experimental evaluation  in Section~\ref{exp} and conclude the paper in Section~\ref{conc}.}

\vspace*{-2.0ex}
\section{Related Work}
\label{related}

In the following, we will review the existing methods on shortest path computation, especially those related to social networks.
Throughout our discussion, we use $n$ and $m$ to denote the number of nodes and edges in the graph $G$, respectively.

\noindent{\bf Online Shortest Path Computation:}
One of the most well-known methods for shortest path computation is Dijkstra's algorithm~\cite{Dijkstra59}.
It computes the single source shortest paths in a weighted graph and can be implemented with $O(m+n\log n)$ time.
If the graph is unweighted (as are many social networks),
a Breadth-First Search (BFS) procedure can compute the shortest path in $O(m+n)$.
However, it is prohibitively expensive to apply these methods to a social network with millions of vertices, even when limiting the search depth to $6$ steps.
First, the average degree in the social network is relatively high. For instance, each user in Facebook on average has about $130$ friends. A straightforward BFS would easily scan one million vertices within $6$ steps.
A simple strategy is to employ bidirectional search to reduce the search space.
Second, due to the existence of hubs and the small-world property, a large number of hubs may be traversed in bidirectional BFS (even within three steps of the start $s$ or end $t$ of the shortest path query).
For instance, in the Orkut graph (a frequently used benchmarking social network), which consists of over $3$ million vertices and $220$ million edges, a bidirectional BFS still needs to access almost $200K$ vertices per query while traditional BFS needs to access almost $1.6$ million vertices per query.

\noindent{\bf Shortest Path Computation on Road Networks:}
Computing shortest path on road networks has been widely studied~\cite{Jing98,Jung02,Shekhar97,Sanders05,BastFSS07,Gutman04,GoldbergSODA05,HananSA08,Sankaranarayanan:2009,Kriegel:2008:HGE,Geisberger:2008:CHF,Tao11,Bauer:2010:CHG,Abraham:2010:HDS,Abraham:2011:HLA}. Here we provide only a short review. A more detailed review on this topic can be found in ~\cite{Delling:2009}.
Several early studies~\cite{Jing98,Jung02,Shekhar97}, such as {\em HEPV}~\cite{Jing98} and {\em HiTi}~\cite{Jung02},  utilize the decomposition of a topological map to speed up shortest path search.
Recently, a variety of techniques~\cite{Delling:2009}, such as $A^\ast$~\cite{GoldbergSODA05},  Arc-flag (directing the search towards the goal)~\cite{Bauer:2010:CHG}, highway hierarchies (building shortcuts to reduce search space)~\cite{Gutman04,Sanders05}, transit node routing (using a small set of vertices to relay the shortest path computation)~\cite{BastFSS07}, and utilizing spatial data structures to aggressively compress the distance matrix~\cite{HananSA08,Sankaranarayanan:2009}, have been developed.
However, the effectiveness of these approaches rely on the essential properties of road networks, such as almost planar, low vertex degree, weighted, spatial, and existence of hierarchical structure~\cite{Gubichev:2010:FAE:1871437.1871503}.
As we mentioned before, social networks have different properties, such as non-spatial, unweighted, scale-free (existence of hubs), and exhibiting small-world properties.
For instance, those techniques utilizing spatial properties (triangle inequality) for pruning the search space immediately become infeasible in social networks.
Also, the high vertex degree (hubs) easily lead to the explosion of the search space.

\noindent{\bf Theoretical Distance Labeling and Landmarking:}
There have been several studies on estimating the distance between any vertices in large (social) networks~\cite{Potamias09,SarmaSMR10,Gubichev:2010:FAE:1871437.1871503, citeulike:7372248,zhao2011,DBLP:conf/icde/Qiao12}.
These methods in general belong to distance-labeling~\cite{DBLP:journals/jal/GavoillePPR04}, which assigns each vertex $u$ a label (for instance, a set of vertices and the distances from $u$ to each of them) and then estimates the shortest path distance between two vertices using the assigned labels.
The seminal work, referred to as the distance oracle~\cite{DBLP:journals/jacm/ThorupZ05},  by Thorup and Zwick shows a $(2k-1)$-multiplicative distance labeling scheme (the approximate distance is no more than $2k-1$ times the exact distance), for each integer $k\geq 1$, with labels of $O(n^{1/k} \log^2n)$ bits.
However, as Potamias {\em et al.}~\cite{Potamias09} argued, for practical purposes,
even $k=2$ is unacceptable (due to the small-world phenomenon).
Recently, Sarma {\em et al.}~\cite{SarmaSMR10} study Thorup and Zwick's distance oracle method on real Web graphs and they find this method can provide fairly accurate estimation.

The pioneering $2$-hop distance method by Cohen {\em et al.}~\cite{cohen2hop} provides exact distance labeling on directed graphs (very similar to the $2$-hop reachability indexing).
Specifically, each vertex $u$ records a list of intermediate vertices $L_{out}(u)$ it can reach along with their (shortest) distances, and a list of intermediate vertices $L_{in}(u)$ which can reach it along with their distances.
To find the distance from $u$ to $v$, the $2$-hop method simply checks all the common intermediate vertices between $L_{out}(u)$ and $L_{in}(v)$ and chooses the vertex $p$, such that $dist(u,p)+dist(p,v)$ is minimized for all $p \in L_{out}(u) \cap L_{in}(v)$. However, the computational cost to construct an optimal $2$-hop labeling is prohibitively expensive~\cite{hopiedbt,DBLP:conf/sigmod/JinXRF09}.

Several works use {\em landmarks} to approximate the shortest path distance ~\cite{Ng01predictinginternet,Kleinberg04,Potamias09,citeulike:7372248,zhao2011,DBLP:conf/icde/Qiao12}.
Here, each vertex precomputes the shortest distance to a set of landmarks and thus the landmark approach can be viewed as a special case of $2$-hop and distance labeling where each vertex can record the distance to different vertices.
Potamias {\em et al.}~\cite{Potamias09} investigate the selection of the optimal set of landmarks to estimate the shortest path distance. Qiao {\em et al.} ~\cite{DBLP:conf/icde/Qiao12} observe that a globally-selected landmark set introduces too much error, especially for some vertex pairs with small distance, and so propose a query-load aware landmark selection method. Zhao {\em et al.}~\cite{zhao2011} introduce Rigel, which utilizes a hyperbolic space embedding on top of the landmark to improve the estimation accuracy.

\noindent{\bf Approximate Shortest Path Computation in Social Networks:}
A few recent studies aim to compute the shortest path in large social networks.  They extend the distance-labeling or the landmarking approach to approximate the shortest paths.
Gubichev {\em et al.}  propose {\em Sketch}, which generalizes the distance oracle method~\cite{DBLP:journals/jacm/ThorupZ05,SarmaSMR10} to discover the shortest path (not only the distance) in large graphs~\cite{Gubichev:2010:FAE:1871437.1871503}.
They observe that the path lengths are small enough to be considered as almost constant and therefore store a set of precomputed shortest path in addition to the distance labeling.
They also propose several improvements, such as {\em cycle elimination} (SketchCE) and {\em tree-based search} (TreeSketch), to boost the shortest path estimation accuracy.
Zhao {\em et al.}~\cite{zhao2011} develop {\em RigelPath} to approximate the shortest path in social networks on top of their distance estimation method, Rigel.
Their basic idea is to use the distance estimation to help determine the search direction and prune search space.
Sketch is the fastest approximate shortest path method, though RigelPath and TreeSketch can be more accurate.
In addition, RigelPath mainly focuses on the undirected graph, while Sketch can handle both directed and undirected graphs.

\noindent{\bf Other Recent Progress on Shortest Path Computation:}
Very recently, there have been a few studies in the database research community on shortest path and distance computation. 
In~\cite{DBLP:conf/sigmod/Wei10}, Wei develops a tree decomposition indexing structure to find the shortest paths in an unweighted undirected graph; In ~\cite{DBLP:conf/sigmod/ChengKCC12}, a hierarchical vertex-cover based approach is developed for single-source on-disk shortest path (distance) computation.
In~\cite{DBLP:journals/pvldb/ChengSCWY12}, Cheng {\em et al.} introduce $k$-reach problem which provides binary answer to whether two vertices are connected by $k$ steps. Also, the $k$-reach indexing approach developed in ~\cite{DBLP:journals/pvldb/ChengSCWY12} is not scalable and can only handle small graphs (as it tries to materializes the vertex pairs within certain distance threshold). 
Finally,  Jin {\em et al.} ~\cite{DBLP:conf/sigmod/JinRXL12} propose a highway-centric labeling (HCL) scheme to efficiently compute distance in sparse graphs. Leveraging highway structure, this distance labeling offers a more compact index size compared to the state-of-the-art 2-hop labeling, and is also able to provide both exact and approximate distance with bounded accuracy. However, it is hard to scale to large social networks as real social networks are generally not sparse and potentially lead to expensive index construction cost and large index size.

\vspace*{-2.0ex}
\section{Hub-Accelerator Framework}
\label{framework}

In this section, we give an overview of the Hub-Accelerator (HA) framework for the shortest path computation.
In the earlier discussion, we observe a love-hate relationship between shortest-path and hubs:  on one hand, any shortest paths likely contain some hubs and thus need to be visited in the shortest path search process; on the other hand, in order to provide the fast shortest path search, we need to try to avoid a full expansion of hub nodes.
We note that in general, the notation of hubs is rather informal though generally based on degree; in this paper, we simply refer to the set of vertices whose degree are the highest (top $\beta$ number of vertices; $\beta$ is a constant and can be specified). 

The design of Hub-Accelerator aims to utilize these hubs for shortest-path computation without fully expanding their neighborhoods. To achieve this,  the following research questions need to answered:  

\noindent{\bf 1. } How we can limit the expansion of hubs during the shortest path search? A hub may have thousands or even millions of connections (neighbors); what neighbors should be considered to be essential and given high priority in the shortest path search? To address this question, we formulate the {\em hub-network} notation, which captures a high-level view of the shortest path and topology between these hubs. The hub-network can be considered a highway structure anchored by hubs for routing the shortest paths in a massive social network. Due to the importance of hubs, most shortest paths between non-hub vertex pairs may need go through such a network, i.e., the starting vertex reaches a hub (as the highway entry), then travels to another hub (as the highway exit), and finally leaves the highway reaching the destination.  In other words, the hub-network can be used to limit (or prioritize) the neighbors of hubs; a hub should only expand within the hub-network. 

\noindent{\bf 2.}  How we can effectively and efficiently utilize the hub-network for shortest path search? Note that the hub-network captures the shortest paths between hubs. However, not all shortest paths between vertices need to go through the hub-network: they may not contain any hub or they may consist of only one hub (in the later case, no traversal may be needed in the hub network). Thus, the problem is how we can extend the typical bidirectional BFS to adopt the hub-network for speeding up the shortest path computation?

\noindent{\bf 3.}
Can we completely avoid the expansion of hubs? In this way, even the hub-network  becomes unnecessary. But what essential information should be precomputed? When the number of hubs is not large,  say $10K$, then the pair-wise distance matrix between hubs may be materialized. For $10K$ hubs, this only costs about $100MB=10K \times 10K b$ (assuming the distance can be held in $8$ bits), but additional memory may be needed to recover the shortest path. Given this, how can bidirectional search take advantage of such a matrix and what other information may also need to be precomputed?

In this work, by investigating and solving these problems, we are able to utilize the hubs effectively to accelerate the shortest path search while significantly reducing or avoiding the cost of expanding them. Specifically, we make the following contributions: 

\noindent{\bf Hub-Network Discovery (Section~\ref{HubNetwork}):}
The concept of hub-network is at the heart of the Hub-Accelerator framework: given a collection of hubs, a {\em distance-preserving subgraph} seeks to extract  a minimal number of additional vertices and edges from the original graphs so that the distance (and shortest path) between hubs can be recovered, i.e., their distances in the hub-network are equivalent to their distances in the original graph. As we mentioned before, the hub-network serves as the highway in the transportation system to enable the acceleration of the shortest path search: any hub will not be fully expanded (in the original graph); instead, only their neighbors in the hub networks will be expanded. Interestingly, though the discovery of a distance-preserving subgraph (and hub-network) seems rather intuitive, the computational aspect of the problem has not been studied before (despite similar notions being defined in theoretical graph theory~\cite{Djokovic1973263}). In Section~\ref{HubNetwork}, we show the NP-hardness of discovering the minimal distance-preserving subgraph and we develop a fast greedy approach to extract the hub-network (and the distance-preserving subgraph).  
Our experimental study shows the degree of hubs in the hub-network is significantly lower than that in the original graph;  thus the hub-network can limit the expansion of hubs and enables faster shortest path computation. 


\noindent{\bf Hub-Network based Bidirectional BFS (Section~\ref{HNBBFS})} 
As we mentioned above, it is nontrivial to incorporate the hub-network into the bi-directional BFS. In general, if we use the hub-network and also expand the hubs within the network, then the searches in both directions cannot simply be stopped when they meet at a common vertex. This is because the hub-network does not capture those shortest paths consisting of only one hub.

\comment{and the search steps may be longer than $k/2$ (which is the stop condition for typical bidirectional BFS). When $k$ is relatively large, say $k \geq 5$, this can potentially introduce a large number of vertices to be visited. To address this problem, in Section~\ref{HubLabeling}, we further introduce a novel {\em Hub-Labeling} technique which utilizes only a small amount of memory to reduce the search step. The basic idea is that for each non-hub vertex in the original graph, it record those hubs within $k$ steps but more than $\lfloor k/2 \rfloor$ steps aways; more importantly, those hubs cannot be reached in a shortest path consisting of other hubs (we refer to them as {\em core-hubs} with respect to a non-hub vertex). In other words, starting from a non-hub vertex, any shortest path  reaching those core-hubs consists of only non-hub vertices. This condition can help significantly prune the number of hubs which need to be recorded. Utilizing such a labeling, we can effectively reduce the steps of the bidirectional BFS to no more than $k/2$ steps for each direction (referred to as the Bounded Bidirectional BFS).}

\noindent{\bf Hub$^2$-Labeling (Section~\ref{Hub2Labeling}):} In this technique, we further push the speed boundary for shortest path computation by completely avoiding expanding any hub. To achieve this, a more expensive though often affordable precomputation and memory cost is used for faster online search. It consists of three basic elements: 1)  First, instead of extracting and searching the hub-network, this technique materializes the distance matrix of those hubs, referred to as the {\em Hub$^2$} matrix. As we mentioned before, even for $10K$ hubs, the matrix can be rather easily materialized. 2)  {\em Hub-Labeling} is introduced so that each vertex will precompute and materialize a small number of hubs (referred to as the core-hubs) which are essential for recovering the shortest path using hubs and hub-matrix.  3) Given the Hub$^2$ distance matrix and hub-labeling, a faster bidirectional BFS can be performed to discover the exact $k$-degree shortest path. It first estimates a distance upper bound using the distance matrix and the hub labeling. No hub needs to be expanded during the bidirectional search, i.e., hub-pruning bidirectional BFS.  



\comment{
\subsection{Our Contribution}
In this paper, a list of novel techniques centered around the hubs (Hub-Acceleration Framework) are introduced to speed up the  shortest path computation. 
Though the shortest path computation has been extensive studied, most of the studies only focus on the approximate distance and shortest path computation on massive social networks~cite{}.  To our best knowledge, this is the first work explicitly addressing the exact distance and shortest path computation in these networks. Furthermore, the techniques based on hubs are also novel and we are not aware any existing research which have investigated the distance-preserving subgraph (hub-network) discovery problem and the explicit labeling method based on the {\em core-hub} concept. Subsequently, the bidirectional search leveraging the hub-network, the core-hub labeling, and the Hub$^2$ distance matrix are also new. 

To summarize, we make the following contributions in this work: }

\comment{
 
Labeling framework and discuss the selection of {\em hub-set}. 
To facilitate our discussion, we introduce the following notation. 
Let $G=(V,E)$ be an undirected graph where $V=\{1,2,...n\}$ is the vertex set and $E \subseteq V \times V$ is the edge set.
The edge from vertex $u$ and $v$ is denoted by $(u,v)$, and we use $P(v_0, v_p) = (v_0, v_1, ..., v_p)$ to denote a simple path between $v_0$ and $v_p$.
The length of a simple path is the number of edges in the path, denoted by $|P(v_0,v_p)|$.
Given two vertices $u$ and $v$, their shortest path $SP(u,v)$ is the path between them with the minimal length.
The distance from vertex $u$ to $v$ is the length of shortest path $SP(u,v)$ between $u$ and $v$, denoted by $d(u,v)$.

To compute the $k$-degree shortest-path in large social network, we develop three new and complementary techniques, {\em Hub$^2$}, {\em Hub-Labeling}, and  {\em Hub-avoidance Bidirectional BFS} (HBBFS). 
They work in harmony and form the Hub$^{2}$-Labeling approach.

\comment{
\begin{example}
Now, let us take the Flicker as an example  network we used in experiments as example to get in-depth understanding.
The network contains around $9.4$ millions vertices and $222$ millions edges and we select $10K$ highest degree vertices as landmarks.
Using core landmark indexing scheme, each vertex only needs to record around $180$ landmarks (i.e., no more than $2\%$ of entire landmarks), and total number of entries for pairwise shortest paths among landmarks is at most $100$ millions.
Distributing materialization cost of shortest paths among landmark to each vertex, each vertex at most needs to record around $190$ ($=180+\frac{100M}{9.4M}$) entries, which is still significantly smaller than $10K$ in traditional landmark indexing scheme.
\end{example}
}

\comment{
Utilizing core landmark indexing, the capacity of $10K$ landmarks is preserved at the expense of index cost equivalent to that using $190$ landmarks.
As we know, the average index cost for each vertex is $\frac{\sum_{u \in V \setminus L} |L_{H}(u)| }{|V|} + \frac{|L|\times |L|}{|V|}$.
Intuitively, as graph size increases, the second term would be much smaller,
and core landmark indexing would bring more benefit regarding index size compared to traditional one.}

\noindent{\bf Advantage over Landmark:} 
In the following, we introduce key observation on the new estimation approach based on the hub-set, hub$^2$, and core-hubs. 
We compare its estimation accuracy with the traditional landmark approach (assuming it can utilize all vertices in the hub-set as landmarks) but with the distance matrix hub$^2$. 
This provides the basis on why we can effectively utilize a large number of landmarks (hub-set vertices) and why our approach can provide a better distance upper-bound for the exact shortest path search in the latter bidirectional BFS. 

\begin{figure}
\centering
\begin{tabular}{c}
\psfig{figure=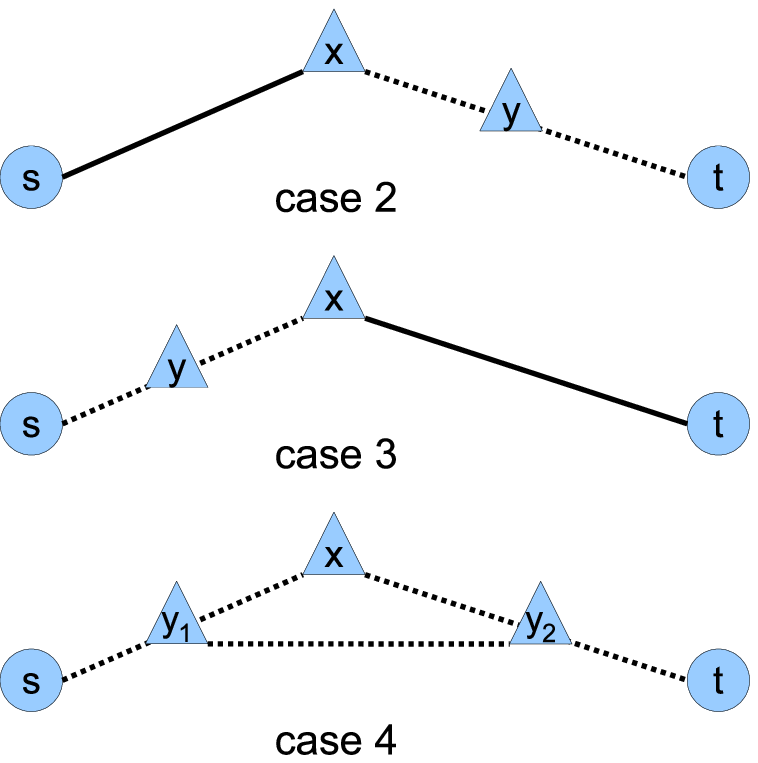,scale=0.65}
\end {tabular}
\caption {Cases in proof of Lemma~\ref{goodcorepath}}
\label{fig:proofcases}
\end{figure}

\blemma
\label{goodcorepath}
Given the hub-set $\mathcal{H} \in V$ in the graph $G=(V,E)$,
the shortest path estimated by using core-hubs together with distance matrix Hub$^2$ is no longer than 
the one obtained by using the traditional landmark scheme (assuming every vertex in $\mathcal{H}$ is a landmark),
i.e., $d_H(s,t|\mathcal{H}) \le d_L(s,t|\mathcal{H})$ for any vertex pair $(s,t)$.
\elemma

\bproof
Without loss of generality, we want to prove that, for any path generated by traditional landmark indexing, we can find one path no longer than that using the core-hubs and distance-matrix. 
For a vertex pair $(s,t)$, let the path obtained by traditional landmark indexing  be $SP(s,x) \circ SP(x,t)$,
where $x \in \mathcal{H}$ is a landmark between $s$ and $t$.
Now, we seek a path based on the core-hubs which is not longer than $SP(s,x) \circ SP(x,t)$ in the following cases:

\noindent 1) $x \in L_{H}(s)$ and $x \in L_{H}(t)$. It is trivial to generate the same path $SP(s,x) \circ SP(x,t)$.

\noindent 2) $x \in L_{H}(s)$ and $x \notin L_{H}(t)$.
According to the definition of {\em core hubs}, there is some landmark $y \in L_{H}(t)$ lying on the shortest path between $x$ and $t$, i.e., $d(x,t)=d(x,y)+d(y,t)$ (case $2$ shown in Figure~\ref{fig:proofcases}).
Since the shortest path $SP(x,y)$ has been materialized in the distance matrix Hub$^2$, we can obtain the path $SP(s,x) \circ SP(x,y) \circ SP(y,t)$ with the same length as $SP(s,x) \circ SP(x,t)$.

\noindent 3) $x \notin L_{H}(s)$ and $x \in L_{H}(t)$. Follow the similar logic in case 2 (case 3 in Figure~\ref{fig:proofcases}).
We guarantee to have $y \in L_{H}(s)$ appearing in $SP(s,x)$.
Thus, we have path $SP(s,y) \circ SP(y,x) \circ SP(x,t)$ which is no longer than $SP(s,x) \circ SP(x,t)$.

\noindent 4) $x \notin L_{H}(s)$ and $x \notin L_{H}(t)$. That is, there are some other core-hubs $y_1 \in L_{H}(s)$ and $y_2 \in L_{H}(t)$ appearing in shortest paths $SP(s,x)$ and $SP(x,t)$, respectively (case 4 shown in Figure~\ref{fig:proofcases}).
According to triangle inequality, we have $d(y_1,x) + d(x,y_2) \ge d(y_1,y_2)$.
Also, shortest path $SP(y_1,y_2)$ has been precomputed and stored in Hub$^2$,
thus we can have the path $SP(s,y_1) \circ SP(y_1,y_2) \circ SP(y_2,t)$ which guarantees to be no longer than $SP(s,x) \circ SP(x,t)$.

Therefore, in any case, the lemma follows.
\eproof

}

\vspace*{-2.0ex}
\section{Hub-Network Discovery}
\label{HubNetwork}

\begin{figure}[t!]
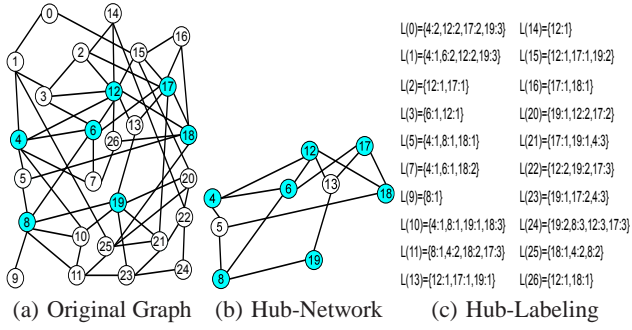

    \label{fig:coreindexexample}
    \centering
    \mbox{
        \subfigure[Original Graph]{\includegraphics[height=1.5in, width=1.0in]{Figures/RunningExampleGraph.epsi}\label{originalgraph}}
        \subfigure[Hub-Network]{\includegraphics[height=0.8in,width=1.0in]{Figures/HubNetwork.epsi}\label{hubnetworkexample}}
        \subfigure[Hub-Labeling]{\includegraphics[height=1.4in,width=1.2in]{Figures/CompleteLabeling.epsi}\label{completelabeling}}
    }
    \vspace*{-2.0ex}
    \caption{{\small Running Example of Hub-Accelerator Framework}}
\label{runningexample}
\vspace*{-3.0ex}
\end{figure}

In this section, we formally define the {\em Hub-Network} (Subsection~\ref{hubnetworkdefinition}) and present an efficient approach to discover it (Subsection~\ref{hubnetworkdiscovery}).

To facilitate our discussion, we first introduce the following notation. 
Let $G=(V,E)$ be a graph where $V=\{1,2,...n\}$ is the vertex set and $E \subseteq V \times V$ is the edge set. 
The edge from vertex $u$ and $v$ is denoted by $(u,v)$, and we use $P(v_0, v_p) = (v_0, v_1, ..., v_p)$ to denote a simple path between $v_0$ and $v_p$.
The length of a simple path is the number of edges in the path, denoted by $|P(v_0,v_p)|$.
Given two vertices $u$ and $v$, their shortest path $SP(u,v)$ is the path between them with the minimal length.
The distance from vertex $u$ to $v$ is the length of shortest path $SP(u,v)$ between $u$ and $v$, denoted by $d(u,v)$.
Note that for a directed graph, the edge set may contain either $(u,v)$, $(v,u)$, or both.
For an undirected graph, the edge has no direction; in other words, it can be considered bidirectional, so the edge set contains either both edges $(u,v)$ and $(v,u)$ or neither of them. 
In undirected graph, the shortest path distance from $u$ to $v$ is equivalent to the one from $v$ to $u$, i.e., $d(u,v)=d(v,u)$.   
The techniques discussed in the paper can be applied both undirected and directed graph; for simplicity, we will focus on the undirected graph and we will briefly mention how each technique can be naturally extended to handle directed graphs. 

\subsection{Distance-Preserving Subgraph and Hub-Network}
\label{hubnetworkdefinition}

Intuitively, a hub-network is a minimal subgraph of the original $G$, such that at least one shortest path between two hubs can be recovered in the subgraph (the distance is preserved). To formally define the hub-network, we first introduce the concept of {\em distance-preserving subgraph} and its discovery. 

\bdefin{\bf Distance-Preserving Subgraph}
Given graph $G=(V,E)$ and a set of vertex pairs $D={(u,v)} \subseteq  V \times V$, a distance-preserving subgraph $G_s=(V_s,E_s)$ of $G$ ($V_s \subseteq V$ and $E_s \subseteq E$) has the following property: for any $(u,v) \in D$, 
$d(u,v|G_s)=d(u,v|G)$, where $d(u,v|G_s)$ and $d(u,v|G)$ are the distances in subgraph $G_s$ and original graph $G$, respectively.   
\edefin

Given a collection of vertex pairs whose distance need to be preserved in the subgraph, the subgraph discovery problem aims to identify  a {\em minimal} subgraph in terms of the number of vertices (or edges).

\bdefin{\bf Minimal Distance-Preserving Subgraph\\ (MDPS) Problem}
Given graph $G=(V,E)$ and a set of vertex pairs $D={(u,v)} \subseteq  V \times V$, the minimal distance-preserving subgraph (MDPS) problem aims to discover a minimal subgraph $G_s^\star=(V_s^\star,E_s^\star)$  with the smallest number of vertices, i.e., $G_s^\star=\arg\min_{|V_s|} G_s$, where $G_s=(V_s,E_s)$ is a distance-preserving subgraph with respect to $D$. 
\edefin

Once all the vertices $V_s^\star$ are discovered, the induced subgraph $G[V^\star_s]$ of $G$ is a candidate minimal subgraph. Note that its edge set may be further sparsified. However, the edge sparsification problem with respect to a collection of vertex pairs (equivalent to the minimal distance-preserving subgraph problem in terms of the number of edges) is equally difficult as the MDPS problem (see discussion below); and the number of edges which can be removed are typically small in the unweighted graph. Thus, we will not explore the further edge reduction in this work. 

Given graph $G=(V,E)$ and a set of hubs $H \subseteq V$, let $D_k$ contain all the hub pairs whose distance is no greater than $k$, then {\em the hub-network is defined as the minimal distance-preserving subgraph of $D_k$ in $G$}. 

\begin{example}
Figure~\ref{runningexample}(a) shows the network we will use as a running example. Figure~\ref{runningexample}(b) is the corresponding hub-network with $H=\{4,6,8,12,17,18,19\}$ (degree $\geq 5$) when $k=4$. Since the pairwise distances between these hubs are all less than $4$, $D_4$ contains all the hub pairs with a total of $15$ vertex pairs. 
\end{example}

\comment{ 
Figure~\ref{runningexample}(b) shows a hub-network of the original graph $G$ (in Figure~\ref{runningexample}(a)) with $H=\{4,6,8,12,17,18,19\}$ (degree $\geq 5$) and for the running example in Figure~\ref{runningexample}, $k=4$ (thus, we focus on the $4$-degree shortest path computation).   
Thus, since the pairwise distance between these hubs is all less than $4$, $D_4$ contain all the hub pairs and there are a total of $15$ vertex pairs.
}

Note that an alternative approach is to build the weighted hub-network which explicitly connects the hub pairs: for instance, if any other hub lies in a shortest paths between two hubs, an edge can be added to directly link them. Indeed, most of the existing studies have adopted a similar approach to build and utilize some {\em highway structure} (but they target mainly road networks, which are rather sparse). However, this approach can lead to a number of problems when searching a massive social network:
1) Such hub-network would be weighted and could be dense (many new edges may need to be added between hubs) and to search through it, Dijkstra's algorithm (or its variant) must be utilized and would be slower than BFS (because of using the priority queue). Higher edge density exacerbates this slowdown.
2) Bidirectional BFS is typically used to search an unweighted network and could be adopted to search the remaining network (excluding the hub-network). However, combining bidirectional BFS with Dijkstra's can be rather difficult;
3) Significant memory may be needed to record such a hub-network as it is rather dense. Moreover, to recover the shortest path, additional information has to be recorded for each added new edge. Considering these issues, we utilize the distance-preserving subgraph as the hub-network, which does not induce additional memory cost, and can naturally support (bidirectional) BFS.  Note that in Sections~\ref{HNBBFS} and ~\ref{Hub2Labeling}, we will study how to use more memory for higher query performance (without involving the difficulty of weighted hub-network). 

To discover the hub-network in a massive social network, we need a fast solution for the Minimal Distance-Preserving Subgraph (MDPS) problem. However, finding the exact optimal solution is hard. 
\bthm
Finding the minimal distance-preserving subgraph of a collection $D$ of vertex pairs in a graph is an NP-hard problem.  
\ethm 
\bproof
We reduce the set-cover decision problem r to the decision version of the minimal distance-preserving subgraph problem. 
In the set-cover decision problem, let $\mathcal{U}$ be the ground set and $\mathcal{C}$ records all the candidate sets, where for any candidate set $C \in \mathcal{C}$ and $C \subseteq \mathcal{U}$.
The set-cover decision problem asks whether there are $K$  or fewer candidate sets in $\mathcal{C}$, such that $\cup_i C_i = \mathcal{U}$. 

Now we construct the following MDPS instance based on a set cover instance: 
consider a tripartite graph $G=(X \cup Y \cup Z, E_{XY} \cup E_{YZ})$ where the vertices in $X$ and $Z$ have one-to-one correspondence to the elements in the ground set $\mathcal{U}$, and the vertices in $Y$ one-to-one correspond to the candidate sets in $\mathcal{C}$. 
For simplicity, let $u \in \mathcal{U} \leftrightarrow x_u \in X (z_u \in Z)$ (vertex $x_u$ ($z_u$) corresponds to element $u$); and let $C \in \mathcal{C} \leftrightarrow y_C \in Y$ (vertex $y_C$ corresponds to candidate set $C$). 
Then, the edge set $E_{XY}$ ($E_{YZ}$) contains all the edges $(x_u,y_C)$ ($(y_C,z_u)$) if element $u$ belongs to the candidate set $C$. 
Note that the tripartite graph can be considered symmetric ($X\equiv Z$ and $E_{XY}\equiv E_{YZ}$). 

We claim that the set-cover decision problem is satisfiable if and only if the following MDPS problem is true: there is a subgraph $G$ with $2|U|+K$ vertices to cover the shortest path distance of $|U|$ vertex pairs ($x_u,z_u$), $u \in U$.

The proof of this claim is as follows. Assume the set-cover problem is satisfiable, let $C_1, \cdots C_k (k \leq K)$ be the $k$ candidate sets which covers the ground set, i.e., $\cup C_i=\mathcal{U}$. Let $Y_C$ include all the vertices in $Y$ corresponding to $C_1,\cdots, C_k$. It is easy to observe the induced subgraph of $G[X \cup Y_C \cup Z]$ can recover the distances of all $|U|$ pairs ($x_u,z_u$), $u \in U$. Note that their  distances in the original graph $G$ and the induced subgraph $G[X \cup Y_C \cup Z]$ are all equal to $2$. 
   
From the other direction, let $G_s$ be the subgraph with $2|U|+K$ vertices which recovers the distances of these $|U|$ pairs. Since the vertices in the pairs have to be included in the subgraph (otherwise, the distance can not be explicitly recovered), the additional $K$ vertices can only come from the vertex set $Y$ (there are $2|U|$ in the vertex pairs from $X$ and $Z$). Note that the distance of ($x_u,z_u$) in the original graph is $2$ and to recover that, a vertex $y_C$ in $Y$ has to appear in the subgraph so that both $(x_u,y_C)$ and $(y_C,z_u)$ are in the subgraph (and in the original graph). This indicates the corresponding candidate set $C$ covers element $u$. Since there are at most $K$ vertices in $Y$, there are at most $K$ candidates needed to cover all the ground set $U$.
\eproof

Based on similar reduction,  we can also prove that finding the minimal distance-preserving subgraph in terms the number of the edges is also an NP-hard problem. Due to simplicity, we will not further explore this alternative in the paper. 

\subsection{Algorithm for Hub-Network Discovery}
\label{hubnetworkdiscovery}

In the subsection, we will discuss an efficient approach for discovering the distance-preserving subgraph and the hub-network. 
To simplify our discussion, we focus on extracting the hub-network, though the approach is directly applicable to any collection of vertex pairs (and thus the general distance-preserving subgraph). 
Recall that in the hub-network discovery problem, given a set $H$ of hubs and a collection $D$ of hub-pairs whose distance is no more than $k$ (for $k$-degree shortest path search), then the goal is to recover the distance for the pairs in $D$  using a minimal (distance-preserving) subgraph. 

To tackle the hub-network (and the distance-preserving subgraph) efficiently, we make the following simple observation. 
For any vertex pairs $(x,y)$ in $D$, if there is another hub $z$, such that $d(x,y)=d(x,z)+d(z,y)$, then we refer to the vertex pair $(x,y)$ as a {\em composite pair}; otherwise, it is a {\em basic pair}, i.e., any shortest path connecting $x$ and $y$ does not contain a hub in $H$. Let $D_b \subseteq D$ be the set of basic pairs. Given this, it is easy to see that {\em if a subgraph can recover all the vertex pairs in $D_b$, then it is a distance-preserving subgraph of $D$ (and thus the hub-network)}. This indicates that we only need to focus on the basic pairs ($D_b$) as the distances of composite pairs can be directly recovered using the paths between basic pairs. 

Considering this, at the high level, the algorithm of the hub-network discovery performs a BFS-type traversal from each hub $h$ and it accomplishes the two tasks:  
1) during the BFS, all basic pairs including $h$, i.e., $(h,v)$, $v \in H$, should be recognized and collected; 
and 2) once a basic pair $(h,v)$ is identified, the algorithm will select a ``good'' shortest path which consists of the minimal number of ``new'' vertices (not included in the hub-network yet). 
In other words, as we traverse the graph from each hub, we gradually augment the hub-network with new vertices to recover the distance (shortest path) of the newly found basic pairs.


\noindent{\bf Recognizing basic pairs:} 
To quickly determine whether the $(h,v)$ is a basic pair during the BFS process starting from hub $h$, we utilize the following observation:   
{\em Let vertex $y$ lie on a shortest path from hub $h$ to non-hub vertex $v$ with distance $d(h,v)-1$ (i.e., $y$ is one hop closer than $v$ with respect to $h$). 
If there is a hub $h^\prime$ appearing in a shortest path from $h$ to $y$ ($h^\prime$ and $y$ may not be distinct),
$h^\prime$ definitely lies on a shortest path from $h$ to $v$ and $(h,v)$ is a composite pair (not basic pair).}
Based on this observation, we simply maintain a binary flag $b(v)$ to denote whether there is another hub appearing in a shortest path between $h$ and $v$.
Specifically, its update rule is as follows: $b(v)=0$ (not basic pair) if $v$ itself is a hub or $b(y)=0$ ($y$ is $v$'s parent in the BFS, i.e., $d(h,y)=d(h,v)-1$ and $d(y,v)=1$).
Thus, during the BFS traversal, when we visit vertex $v$, if its flag $b(v) = 1$ (true) meaning there is no other hubs lying on the shortest path between $h$ and $v$ and we are able to recognize it is a basic pair.

\noindent{\bf Selecting a ``good'' shortest path between basic pairs:}  
To select a good shortest path between basic pairs $h$ and $v$, a basic measurement is the number of ``new vertices'' that need to be added to the hub-network. As a greedy criterion, the fewer that need to be added, the better is the path. 
To compute this, for any shortest path from starting point $h$ to $v$, a score $f$ records the maximal number of vertices which are already in the hub-network. This measure can be easily  maintained incrementally. Simply speaking, its update rule is as follows: $f(v)=max f(u)+1$ if $v$ itself is in the hub-network or $f(v)= max f(u)$, where $u$ is $v$'s parent in the BFS (a shortest path from $h$ to $v$ go through $u$ and $u$ directly links to $v$). Also vertex $v$ records $u$ which has the maximal $f$ for tracking such a shortest path (with maximal number of vertices in the hub-network).  
Finally, we note that only for vertices $v$ with $b(v)=1$, i.e., when the shortest path between $h$ and $v$ does not go through any other hub, does a score $f$ need to be maintained. Otherwise, $v$ and its descendents cannot produce any basic pairs. 

\begin{algorithm}
{\small
\caption{BFSExtraction($G=(V,E)$, $h$, $H$, $H^\star$)}
\label{alg:bfsextraction}
\begin{algorithmic}[1]
\STATE Initialize $b(u)\leftarrow 1$; $f(u) \leftarrow 0$ for each vertex $u$;
\STATE $level(h) \leftarrow 0$; $Q \leftarrow \{h\}$ \COMMENT{queue for BFS};
\WHILE{$Q \neq \emptyset$ \COMMENT{vertices with $b(u)=0$ visited first at each level}}
    \STATE $u \leftarrow Q.pop()$;
    \IF{$u  \in \mathcal{H}$ and $level(u) \ge 1$ and $b(u)=1$ \COMMENT{basic pair}}
        \STATE extract shortest path $SP(h,u)$ with minimal $f(u)$ and add to $H^\star$ 
        \STATE $b(u) \leftarrow 0$ \COMMENT{all later extension will become false}
    \ENDIF
    \IF {$level(u)=k$ \COMMENT{no expansion more than level$k$ for $k$-degree shortest path}} 
        \STATE continue; 
    \ENDIF 
    \IF{$b(u)=1$ and $u \in H^\star$}
            \STATE $f(u) \leftarrow f(u) + 1$ \COMMENT{increase $f$}    
    \ENDIF
    \FORALL{$v \in neighbor(u)$ \COMMENT{$(u,v) \in E$; expanding $u$}}
            \IF{$v$ is not visited}
                \STATE add $v$ to queue $Q$;
            \ELSIF{$level(v) = level(u)+1$} 
                \IF {$b(u)=0$ \COMMENT{update $b$}}
                    \STATE $b(v) \leftarrow 0$;
            	 \ELSIF{$b(v)=1$ and $f(u)>f(v)$ \COMMENT{update $f$}}
                    \STATE $f(v) \leftarrow f(u)$ and $parent(v) \leftarrow u$;
                \ENDIF
			\ENDIF
    \ENDFOR
\ENDWHILE
\end{algorithmic}
}
\end{algorithm}

\noindent{\bf Overall Algorithm:}
The outline of this BFS-based procedure for discovering the hub-network is described in Algorithm~\ref{alg:bfsextraction}.
Here $H^\star$ is the set recording the vertices in the hub-network. Initially, $H^\star=H$ and then new vertices will be added during the processing. 
Note that in the queue for BFS traversal (Line $3$), we always visit those vertices with $b(u)=0$, i.e., they and any of their descendents (in the BFS traversal) will not form a basic pair, and thus the score $f$ does not need to be maintained for them. 
Once a hub is visited and it initially has $b(u)=1$, then $(h,u)$ is a basic pair (Line $5$); we will extract the shortest path which has the maximal number of vertices in the hub-network  and add the new vertices to $H^\star$ (Line $6$). 
Now, since the descendent of this hub (in the BFS traversal) will not form a basic pair, we simply change its flag to false, i.e., $b(u)=0$ (Line $7$). 
Also, since we are only interested in the shortest path within $k$-hop, we will not expand any vertex with distance to $h$ to be $k$ (Lines $9-11$). 
Before we expand the neighbors of $u$, we also need to update its $f$ score based on whether $u$ itself is in the hub-network (Line $12-14$). 

The complete expansion of a vertex $u$ is from Line $15$ to $28$. We will visit each of its neighbors $v$. 
If $v$ has not been visited, we will add it to the queue for future visiting (Line $16-18$). 
Then we perform the incremental update of flag $b(v)$ and score $f(v)$. 
Flag $b(v)$ will be turned off if $b(u)=0$ (Line $20-22$) and if $f(u)$ is larger than $f(v)$, i.e., the shortest path from $h$ to $u$ has the largest number of vertices so far in the hub-network. Vertex $v$ will record $u$ as the parent (for shortest path tracking) and $f(v)$ is updated (Line $24-26$). 
This procedure will be invoked for each hub in $H$.

\begin{figure}
\centering
\begin{tabular}{c}
\psfig{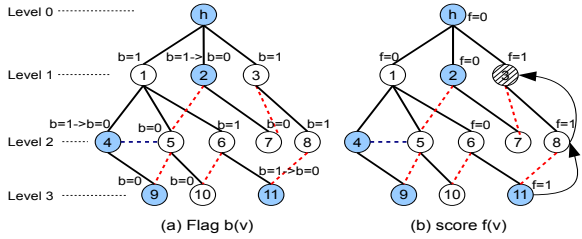}
\end {tabular}
\caption{Incremental Maintenance of flag $b$ and score $f$}
\label{fig:bfsexample}
\end{figure}

\begin{example}
Figure~\ref{fig:bfsexample} illustrates the flag $b$ and score $f$ in the BFS process. 
Here the vertices $h$, $2$, $4$, $9$, and $11$ are hubs. 
In Figure~\ref{fig:bfsexample} (a), $(h,2)$, $(h,4)$, and $(h,11)$ are basic pairs; the flag $b$ changes from $b=1$ originally to $b=0$ (Lines $5$-$7$). 
After the flag $b$ of $2$,$4$, and $11$ changes to false ($b=0)$, all their descendents in the BFS traversal become false. 
For instance, the flag $b$ of vertex $5$ is false as it is also considered hub $2$'s descendent. 
In Figure~\ref{fig:bfsexample}(b), the shaded vertex $3$ indicates it is already included in the hub-network ($3 \in H^\star$). Therefore, vertex $11$ points to vertex $8$ (parent(11)=8 and parent(8)=3) as its $f$ score is higher than the that of vertex $6$. 
\end{example}

\bthm
If we invoke Algorithm~\ref{alg:bfsextraction} for each $h\in H$, then the induced subgraph $G[H^\star]$ is a hub-network of $H$ with respect to the $k$-degree shortest path.  
\ethm
\bproof
The correctness of the algorithm can be derived from the following two observations: 
1) for any basic pair $(h,u)$ with distance no more than $k$, there is at least one shortest path in $G[H^\star]$ as the algorithm explicitly extracts a shortest path and adds all its vertices to $H^\star$; 
2) for any composite pair $(h,h^\prime)$ with distance no more than $k$, then it can always be represented as a sequence of basic pairs, which has at least one shortest path in $G[H^\star]$. 
Thus, for any hub pair $(h,h^\prime)$ with distance no more than $k$, their distance (at least one shortest path) is preserved in the induced subgraph $G[H^\star]$. 
\eproof

The computational complexity for hub-network discovery as described in Algorithm~\ref{alg:bfsextraction} is basically equivalent to that of a simple BFS procedure.  
The overall procedure takes $O(\sum_{h \in H} (|N_k(h)|$$+|E_k(h)|))$ time, where $H$ is the hub set, and $N_k(h)$ and $E_k(h)$ are the number of vertices and edges, respectively, in $u$'s k-degree neighborhood.
We also note that this algorithm works correctly for both undirected and directed graphs.   
Interestingly, we note the following property of applying Algorithm~\ref{alg:bfsextraction} for an undirected graph. 

\blemma
\label{symmetric}
Let $(u,v)$ be a basic hub pair in an undirected graph. Consider Algorithm~\ref{alg:bfsextraction} performs BFS from $u$ first and it discovers the shortest path $SP(u,v)$. When it performs BFS from $v$ and discovers the symmetric basic pair $(v,u)$, the algorithm will not add any additional new vertices. 
\elemma
\bproof
The score $f$ guarantees $f(v)=|SP(v,u)|$=$|SP(u,v)|$ and thus a shortest path as ``good'' as $SP(u,v)$ will be extracted which does not need to add any new vertices to $H^\star$.  
\eproof

This observation leads to the simple bound constraint of the hub-network (the final size of $H^\star$) and the result of Algorithm ~\ref{alg:bfsextraction} will match such a bound. 

\blemma
Let $D_k^b \subseteq D_k \subseteq H \times H$ be the set of all unique basic hub pairs whose distance is no more than $k$, then, 
\[ |H^\star| \leq \sum_{(u,v) \in D_k^b} (d(u,v)-1)+|H| \leq \frac{|H| B}{2} (k-1) +|H|, \] where $B$ is the average numnber  of basic pairs per hub. 
\elemma
\bproof
The term $\sum_{(u,v) \in D_k^b} (d(u,v)-1)$ corresponds to the definition that any basic pair needs to recover only one shortest path; this also  corresponds to the worst case scenario in Algorithm~\ref{alg:bfsextraction},  where for any basic pair, all non-hub vertices along a new shortest path need to be added to $H^\star$. Note that for undirected graph $D_k^b$ treats basic pairs $(u,v)$ and $(v,u)$ as a single one. 
This directly leads to the term $|H| B/2 (k-1)$, which contemplates the maximal distance between any basic hub pair is $k$ and only one shortest path needs to be recovered for symmetric basic pairs $(u,v)$ and $(v,u)$. 
 Algorithm~\ref{alg:bfsextraction} also holds that (Lemma~\ref{symmetric}).  
Note that the result holds for directed graph as well where $B$ is the total degree of both incoming and outgoing edges. 
\eproof

\comment{ 
\vspace*{2.0ex}
\noindent{\bf Improving the $f$ score:} 
The $f$ score plays an important role in the greedy strategy: through selecting a good shortest path with a maximal $f$, the existing vertices in $H^\star$ can be reused. 
However, when some new vertices added into $H^\star$, the current $f$ score may not be able to immediately reflect that as the $f$ score is incrementally maintained. Suppose we have two basic pairs $(h,u)$ and $(h,v)$ and $u$ and $v$ are at the same level during the BFS traversal. When we add some new vertices from $SP(h,u)$ to $H^\star$, the $f$ score for $v$ has been computed and the new vertices will not affect the decision for $v$. To take such immediate benefit into consideration is too costly as we have to recompute $f$ for $v$.  
Furthermore, in the first several BFS process, very few vertices have been  added in $H^\star$ and any shortest path may have the same score. Thus, $f$ score is not effective in the initial selection of good shortest path and we also need a mechanism to select a ``good'' shortest path when $f$ score is tied. 

Considering these factors, we can utilize some additional measurement, which can be easily computed and complementary to the $f$-score. Basically, each vertex is assigned to an extra score to indicate its importance. Such a score servers for two purpose: 1) when two paths have the same $f$-score, the new score on the paths can be used to break a tie; 2) even the $f$-score is not immediately updated after some vertices being added to $H^\star$, the new score help construct certain consensus on what vertices tend to be selected (even without one explicitly informing the other). 

So, what score can serve for such a purpose? Intuitively a vertex connects more pairs of basic pairs with shortest path, it should more likely to be included in the hub-network. Given this, we introduce the following {\em H-betweenness} score: 

\[B_H(v)=|\{(h,h^\prime): d(h,h^\prime)=d(h,v)+d(v,h^\prime) \wedge d(h,h^\prime)\leq k\}| \] 
Basically, $B_H(v)$ is the number of basic hub pairs (with $k$ hops) which has a shortest path passing through $v$. 
Given this, when two shortest path ties at $f$-score, we can select the one with higher total $B_H(v)$ for all the vertices on the path (or those not in $H^\star$). 
To compute such a score, a BFS procedure similar to Algorithm~\ref{alg:bfsextraction} needs to be invoked for each hub. Specifically, when a basic pair is recognized, each intermediate non-hub vertices increase their count for $B_H(v)$. 
The preprocessing step needs to be performed before the actual hub-network discovery and thus for each vertex, two BFS process needs to be performed. 
Can we avoid such additional BFS traversals for computing H-betweenness? 
A simple surrogate is to simply use the vertex degree: the higher the degree, the more likely it can contribute to shortest paths. 
In Section~\ref{experiment}, we will study these different alternatives to improve score $f$ for extracting the hub-network. 
Finally, we note that the H-betweenness can be considered as a variant of the classical {\em vertex betweenness}~\cite{citeulike:1025135}, which measures the vertex importance in terms of all shortest paths in the network, and its computation complexity $O(|V||E|)$~\cite{Brandes2001} is prohibitively expensive for large social graphs.}

\vspace*{-2.0ex}
\section{Hub-Network based Search}
\label{HNBBFS}

In this section, we describe the hub-network based bidirectional BFS. 
The main challenge here is given a hub-network, how we can leverage it to maximally reduce the expansion of hubs and still guarantee to discover the correct $k$-degree shortest path? 
Recall that a key reason for introducing the hub-network is to use it to constraint the expansion of hubs. 
Thus, a basic search principle is that {\em any hub will only visit its neighbors in the hub-network}. 
But what about any non-hub vertices $v$ in the hub-network, such as $v \in H^\star \setminus H$? Should they be expanded only within the hub-network or should they be treated as the remaining vertices outside the hub-network?
Furthermore, in traditional bidirectional BFS, when two searches (forward and backward) meet for the first time, the shortest path is discovered. Unfortunately, this does not necessarily hold if the hub is not fully expanded and thus the question becomes: what should be the correct stop condition? The stop condition is crucial as it determines the search space and the correctness of discovering the exact shortest path. 

In the following, we first describe the hub-network based bidirectional BFS algorithm (Subsection~\ref{algorithmdescription}) and then we prove its correctness and discuss its search cost (Subsection~\ref{discussion}).

\vspace*{-1.0ex}
\subsection{HN-BBFS Algorithm}
\label{algorithmdescription} 

The Hub-Network based Bidirectional BFS (HN-BBFS) algorithm consists of a two-step process: 
1) ({\bf Meeting step}) A bidirectional search will traverse both hub-network and remaining graphs until the forward and backward searches meet at the first common vertex; 
2) ({\bf Verification step}) Next, the searches continues in the remaining graphs (not hub-network) to verify whether the path discovered in the first step is shortest. If not, this step will discover an alternative shortest path.

\noindent{\bf Expansion Rule:}
In the Meeting step, the forward (backward) BFS follows the following rules to expand vertex $v$ in $G$: 
1) if a vertex is a hub, then it only expands its neighbors in the hub-network; 
2) if a vertex is a regular vertex (not in the hub-network), then it expands all its neighbors; 
3) for a vertex is a non-hub vertex but in the hub-network, $H^\star \setminus H$, if the BFS traversal first reaches it through a hub, then it only expands its neighbors in the hub-network; otherwise, it is considered a regular vertex (no shortest path from start (end) vertex to it going through a hub).  
In the Verification step, both forward and backward BFS traversals will continue but they will not need to expand any hub, and any regular vertex and non-hub vertices in the hub-network will expand all their neighbors in the entire network. 

\noindent{\bf Stop Condition:} The stop condition for the forward (backward) BFS in the Verification step is as follows. 
Let $dist$ be the shortest path distance discovered so far; let $d_s^h$ ($d_t^h$) be the distance between $s$ ($h$) to its closest hub $h$; let $level_f$ ($level_b$) be the current level being traversed by forward (backward) BFS. 
Then, the forward (backward) BFS will stop when the following condition is met: 
\beqnarr
dist \geq level_f+d_s^h+ 1 \ \ (dist \geq level_b+d_t^h+1) 
\label{stopcondition}
\eeqnarr

\begin{algorithm}
\caption{HN-BBFS($G$, $G[H^\star]$, $s$, $t$)}
\label{alg:step2query}
\begin{small}
\begin{algorithmic}[1]
\STATE $Q_f \leftarrow \{s\}$; $Q_b \leftarrow \{t\}$; \COMMENT{Queues for forward and backward search}
\STATE $dist \leftarrow k+1$; $met \leftarrow false$; 
\WHILE{($Q_f \neq \emptyset$ AND $Q_b \neq \emptyset$) AND NOT $met$ AND $d(s,Q_f.top)+d(Q_b.top,t)<dist$} 
        \STATE ForwardSearch$(Q_f, false)$;  \COMMENT{not Verification Step}
        \STATE BackwardSearch$(Q_b, false)$; 
\ENDWHILE
\STATE $stop_f \leftarrow false$; $stop_b \leftarrow false$;
\WHILE{(NOT (($Q_f = \emptyset$ OR $stop_f$) AND ($Q_b = \emptyset$ OR $stop_b$)))}
           \STATE NOT $stop_f$: ForwardSearch$(Q_f,true)$; \COMMENT{true: Verification Step}
           \STATE NOT $stop_b$: BackwardSearch$(Q_b,true)$ 
\ENDWHILE
\RETURN {$dist$ and shortest path};
\PROCEDURE{ForwardSearch($Q_f$,$Verification$)}
\STATE $u \leftarrow Q_f.pop()$ \COMMENT{if $Verification$ is true, only out-hub-network vertices will be visited}
\STATE $u$ is set to be visited by forward search;
\FORALL{$v \leftarrow neighbor(u)$ \COMMENT{if $u$ is a hub or there is a shortest path from $s$ to $u$ via a hub, $neighbor(u)$ is within the hub-network}}
    \IF{$v$ is visited by backward search \COMMENT{searches meet}}
        \IF{$d(s,u)+d(v,t)+1 < dist$}
            \STATE update $dist$ and the shortest path correspondingly;
            \IF {NOT $met$ \COMMENT{the first time meet}}
                 \STATE $met \leftarrow true$ 
            \ENDIF
        \ENDIF
    \ENDIF
    \IF{$v$ is not visited AND NOT ($Verification$ and $v \in H$)}
        \STATE $Q_f.push\_back(v)$;
    \ENDIF
    \IF {$Verification$ AND $dist \geq d(s,v)+d_t^h+1$} 
         \STATE $stop_f \leftarrow true$;
    \ENDIF
\ENDFOR
\end{algorithmic}
\end{small}
\end{algorithm}

\comment{
\STATE $Q \leftarrow Q_f$; $Q_f \leftarrow \emptyset$; $r \leftarrow \infty$;
\STATE $mp \leftarrow -1$; \COMMENT{$mp$ is the meeting point between forward search and backward search}
\WHILE{$Q \neq \emptyset$}
    \STATE $u \leftarrow Q.pop()$;
    \FORALL{$v \leftarrow Neighbor(u)$}
        \IF{$v$ is visited in backward search}
            \IF{$l_f(u)+l_b(v)+1 < r$}
                \STATE $r \leftarrow l_f(v)+l_b(v)$;
                \STATE $mp \leftarrow v$;
            \ENDIF
        \ENDIF
        \IF{$v$ is not visited and $l_f(v) \le 3$}
            \IF{$v$ is gate vertex}
                \STATE $Q_f.push\_back(v)$;
            \ELSIF{$l_f(v)<3$}
                \STATE $Q.push\_back(v)$;
            \ENDIF
            \STATE $Prev_f[v] \leftarrow u$;
        \ENDIF
    \ENDFOR
\ENDWHILE
\RETURN $(r,mp)$;

\COMMENT{Step 1: local landmark-based approximation}
\STATE $dist \leftarrow \infty$; $P \leftarrow \emptyset$; \COMMENT{$P$ is the shortest path so far}
\FORALL{$x \in L^e_{out}(s)$}
    \FORALL{$y \in L^e_{in}(t)$}
        \IF{$d(s,x)+d(x,y)+d(y,t) < dist$}
            \STATE $dist \leftarrow d(s,x)+d(x,y)+d(y,t)$;
            \STATE $P \leftarrow SP(s,x) \circ SP(x,y) \circ SP(y,t)$;
        \ENDIF
    \ENDFOR
\ENDFOR \\

\COMMENT{Step 2: modified bidirectional BFS}
\STATE $forward \leftarrow true$; \COMMENT{flag used to indicate the search direction}
\STATE $Q_f \leftarrow \{s\}$; $Q_b \leftarrow \{t\}$; \COMMENT{$Q_f$ and $Q_b$ are queues for forward and backward search}
\WHILE{$Q_f \neq \emptyset$ or $Q_b \neq \emptyset$}
    \IF{$forward = true$}
        \STATE $(d_f, P_f) \leftarrow ForwardSearch(Q_f)$; \COMMENT{$d_f$ and $P_f$ are best distance and shortest path obtained from current forward search}
        \IF{$d_f < dist$}
            \STATE $dist \leftarrow d_f$;
            \STATE $P \leftarrow P_f$;
        \ENDIF
    \ELSE
        \STATE $(d_b, P_b) \leftarrow BackwardSearch(Q_b)$;
        \IF{$d_b < dist$}
            \STATE $dist \leftarrow d_b$;
            \STATE $P \leftarrow P_b$;
        \ENDIF
    \ENDIF
    \IF{$d(s,Q_f.top) + d(Q_b.top,t) \ge dist$ \COMMENT{$Q_f.top$ and $Q_b.top$ are top elements in both queues}}
        \STATE BREAK;
    \ENDIF
\ENDWHILE
\end{algorithmic}
\end{small}
\end{algorithm}

\STATE $Q_f \leftarrow \{s\}$; $Q_b \leftarrow \{t\}$;
\COMMENT{$Q_f$ and $Q_b$ are queues for forward search and backward search, respectively}
\STATE $d \leftarrow \infty$;\\
\COMMENT{perform forward search and backward search}
\WHILE{$Q_f \neq \emptyset$ or $Q_b \neq \emptyset$}
    \STATE $(r,mp) \leftarrow ForwardSearch(Q_f)$;
    \IF{$r < d$}
        \STATE $d \leftarrow r$;
    \ENDIF
    \STATE $(r,mp) \leftarrow BackwardSearch(Q_b)$;
    \IF{$r < d$}
        \STATE $d \leftarrow r$;
    \ENDIF
\ENDWHILE \\
\COMMENT{backtrace shortest path}
\STATE $SP \leftarrow \emptyset$;
\STATE $u \leftarrow mp$;
\WHILE{$Prev_f[u] \neq \emptyset$}
    \STATE $SP.push\_back(u)$;
    \STATE $u \leftarrow Prev_f[u]$;
\ENDWHILE
\STATE $u \leftarrow Prev_b[mp]$;
\WHILE{$Prev_b[u] \neq \emptyset$}
    \STATE $SP.push\_front(u)$;
    \STATE $u \leftarrow Prev_b[u]$;
\ENDWHILE
\RETURN $(SP,d)$;

\begin{algorithmic}[1]
\PROCEDURE{ForwardSearch($Q_f$)}
\STATE $Q \leftarrow Q_f$; $Q_f \leftarrow \emptyset$; $r \leftarrow \infty$;
\STATE $mp \leftarrow -1$; \COMMENT{$mp$ is the meeting point between forward search and backward search}
\WHILE{$Q \neq \emptyset$}
    \STATE $u \leftarrow Q.pop()$;
    \FORALL{$v \leftarrow Neighbor(u)$}
        \IF{$v$ is visited in backward search}
            \IF{$l_f(u)+l_b(v)+1 < r$}
                \STATE $r \leftarrow l_f(v)+l_b(v)$;
                \STATE $mp \leftarrow v$;
            \ENDIF
        \ENDIF
        \IF{$v$ is not visited and $l_f(v) \le 3$}
            \IF{$v$ is gate vertex}
                \STATE $Q_f.push\_back(v)$;
            \ELSIF{$l_f(v)<3$}
                \STATE $Q.push\_back(v)$;
            \ENDIF
            \STATE $Prev_f[v] \leftarrow u$;
        \ENDIF
    \ENDFOR
\ENDWHILE
\RETURN $(r,mp)$;
\end{algorithmic}

}

\noindent{\bf Overall Algorithm:}
Hub-Network based Bidirectional BFS (HN-BBFS) is sketched in Algorithm~\ref{alg:step2query}. 
Note that {\em BackwardSearch} is essentially the same as {\em ForwardSearch} and is omitted for simplicity.
Initially, $dist$ is set to be $k+1$ for $k$-degree shortest path search (indicating no path within $k$-hops) and the $met$ condition is false (Line $2$). 

The first step (Meeting Step) is carried out by the first while loop (Lines $3-6$), 
where a forward search and a backward search are employed in an alternating manner. 
In {\em ForwardSearch} (and {\em BackwardSearch}), 
a vertex in the corresponding queue $Q_f$ ($Q_b$) is expanded.
The expansion rule as described earlier is used in Line $15$. Basically, if a vertex is a hub or is in the hub-network, $H^\star \setminus H$, but the BFS traversal first reaches it through a hub (there is a shortest path from $s$ to $u$ via a hub), it is considered ``in-hub-network''. Otherwise, it is ``out-hub-network''. For an in-hub-network vertex, BFS only expands its neighbors in the hub-network. Note that recognizing these ``in-hub-network'' vertices is straightforward and can be incrementally computed (similar to using the flag $b$ in Algorithm~\ref{alg:bfsextraction}). 
Once a forward (backward) search visits a vertex already visited by the backward (forward) search, a candidate shortest path is discovered and $met$ is set to be true. 
Note that when $Verification$ is false (at the first step), every vertex (both hubs and non-hubs) will be visited and expanded. 

Once $met$ turns true, the second step (Verification Step) is carried out by the second while loop (Lines $8-11$). 
Before the forward stop condition is met ($stop_f$ is false), the ForwardSearch will continue. 
However, only out-hub-network vertices will be visited and expanded (Line $13$ and Lines $24-26$).
Also, during the expansion process, the candidate shortest path can be updated (Lines $17-19$). 
Finally, when the stop condition is met (Line $26$: $d(s,v)$ is the current BFS level being expanded, thus $level_f$), $stop_f$ will become true and no forward search will not performed (Line $9$). 
Note that $d_s^h$ ($d_t^h$) can be easily computed during the BFS traversal: the first time a hub is visited, its distance to $s$ is recorded as $d_s^h$.

\vspace*{-1.0ex}
\subsection{Correctness and Search Cost}
\label{discussion}
We now discuss the correctness of HN-BBFS (Algorithm~\ref{alg:step2query}) and then its search cost (especially in terms of the new Stop condition, Formula~\ref{stopcondition}). To prove the correctness of HN-BBFS, we will make the following important observations: 

\blemma
\label{hubdistance}
For any hub $h \in H$, during the first step (Meeting Step), the distance $d(s,h)$ computed using the forward BFS search, i.e., the
number of traversal levels to reach $h$, is the exact shortest path distance between $s$ and $h$. 
The same holds for $d(h,t)$ for the backward BFS traversal. 
\elemma
\bproof
If $s$ is a hub, then based on the hub-network definition, this clearly holds. 
If $s$ is not a hub, then one of the following two cases must hold: 1) All shortest paths between $(s,h)$ do not contain a hub except $h$, so the forward BFS finds the shortest path distance $d(s,h)$ by traversing only non-hub vertices in the original graph; 2) There is a shortest path between $(s,h)$ containing another hub, so there is always $h^\prime$, such that $(s,h^\prime)$ does not contain any hubs and $(h^\prime,h)$ can be discovered in the hub-network.  
\eproof

Lemma~\ref{hubdistance} demonstrates the power of the hub-network and shows that HN-BBFS can correctly calculate the shortest path (distance) between query vertices to hubs (and between hubs). However, despite this, the candidate shortest path being discovered at the first meeting vertex may not be the exact one. The following lemma categorizes the exact shortest paths if they are shorter than  the candidate shortest path discovered in the first step (Meeting Step). 

\blemma
\label{exactshortestpath}
Assuming $u$ is the meeting vertex where forward and backward search first meet (Lines $22-26$ in Algorithm~\ref{alg:step2query}), the candidate shortest path is denoted as $SP(s,u,t)$ and the distance $dist$ is $d(s,u)+d(u,t)$. If there is a shorter path, then it must contain a hub $h$, such that the exact shortest path can be represented as two segments $SP(s,h)$ and $SP(h,t)$.
Moreover, either 1) $SP(s,h)$ contains no hub other than $h$ with distances $d(s,h) \geq d(s,u)$ and $d(h,t)<d(u,t)$, or 2) $SP(h,t)$ contains no hub other than $h$ with distances $d(s,h)<d(u,t)$ and $d(h,t)\geq d(u,t)$.
\elemma
\bproof
We prove this by way to contradiction. It the lemma does not hold, then the following two types of paths cannot be shorter than the discovered candidate shortest path: 1) there is no hub in the exact shortest path $SP(s,t)$, and 2) there are two hubs $h_s$ and $h_t$, such that the shortest path has three segments: $SP(s,h_s)$, $SP(h_s,h_t)$ and $SP(h_t,t)$ where $d(s,h_s) < d(s,u)$ and $d(h_t,t) < d(u,t)$.  
For the first case, the bidirectional BFS should be able to find such a path (if they are shorter than the candidate $SP(s,u,t)$) earlier as it only involves visiting non-hub vertices in the graph. 
For the second case, based on Lemma~\ref{hubdistance},  Algorithm~\ref{alg:step2query} computes the exact $d(s,h_s)$ and $d(h_t,t)$ before the two BFS met at $u$ and the hub-network encodes the correct distance between $d(h_s,h_t)$. 
Thus, if $d(s,h_s)+d(h_s,h_t)+d(h_t,t)< d(s,u)+d(u,t)$, this shortest path should be discovered (met at an in-hub-network vertex) during the first step (Meeting Step).  
Since both cases are impossible, the lemma holds. 
\eproof

\bthm
\label{correctness}
The Hub-Network based Bidirectional BFS approach (HN-BBFS, Algorithm~\ref{alg:step2query}) guarantees the discovery of the exact $k$-degree shortest path.
\ethm
\bproof
Basically, we need show that when the stop condition is met, no shorter alternative paths exists. 
By Lemma~\ref{exactshortestpath}, if a shortest path exists that is better than the candidate shortest path $SP(s,u,t)$), it must follow one of two simple formats. 
These formats suggest we only need to extend out-hub-network vertices until they meet a hub already visited from the other direction ($d(s,h_s)<d(s,u)$ or $d(h_t,t)<d(u,t)$).
If such a path can be found, it must be shorter than the already discovered distance $dist$, i.e., $dist>level_f+1+d_t^h$ (the best case situation is when the shortest path extends from the current step by one step to a hub closest to the query vertices). Clearly, if this does not hold, any shortest path in this format will not be smaller than $dist$. 
\eproof

In classical Bidirectional search, once both directions meet at a common vertex, the search can be stopped and the exact shortest path is discovered. 
However, in HN-BBFS, in order to reduce the expansion  of hubs, some additional traversal (Verification Step) has to be taken. Clearly, if we need to walk $k/2$ additional steps, then the benefit of HN-BBFS can be greatly compromised.  

So, what is the average number of steps HN-BBFS needs to take for a typical (random) query in the Verification step? The number is close to $zero$ or at most one. To illustrate, first consider the distance between two vertex pairs to be $6$ (since most distances are less than that in social networks~\cite{sysomos}), and assume $s$ and $t$ are not hubs (because there are few hubs) but each of them has a direct hub-neighbor $d_s^h=1$ ($d_t^h=1$). Note that both directions typically traverse at most three steps, i.e., $level_f=level_b=3$. Thus, at most one extra step needs to be taken in this case to make the stop condition true: $dist-level_f-d_t^h-1 \geq 0$, where $level_f=4$. Similarly, let us consider the distance to be $4$ and assume each direction has taken $2$ steps in the Meeting Step. In this case, there is no need to take an additional step (assuming $s$ and $t$ are not hubs), and we can immediately recognize that the candidate shortest path is indeed the exact one.  Finally, we note that when $dist-level_f-d_t^h-1=1$, i.e., the last step of BFS for Verification, there is no need to expand all the neighbors of a given vertex. Only its immediate hub-neighbors need to be expanded and checked (Lemma~\ref{exactshortestpath} and Theorem~\ref{correctness}). To facilitate this, the neighbors of regular vertices can be reorganized so that the hub-neighbors and non-hub-neighbors are separately recorded.

\comment{
In this section, we will discuss how to construct the hub-set $\mathcal{H}$ (Subsection~\ref{hubsetselection}), and how to compute the core-hub labeling $L_H$ for each vertex $v$ efficiently (Subsection~\ref{indexingalgo}). 

\subsection{Hub-Set Construction}
\label{hubsetselection}

The vertices in hub-set have similar functionality to the landmarks, which aim to maximally {\em cover} the shortest paths in the graph. 
Intuitively, a  shortest path is covered by the hub-set if it consists of at least one vertex in the hub-set. 
Formally, we define the goodness of a hub-set using the {\em coverage} criterion: 

\bdefin{\bf (Hub-Set Coverage)}
Given a hub-set $\mathcal{H}$ and vertex pair $(s,t)$, if the estimated shortest distance using hub-set indexing scheme is {\em exact} distance (i.e., $d(s,t)=$ \\ $min_{x \in L_{H}(s) \wedge y \in L_{H}(t)} (d(s,x)+d(x,y)+d(y,t))$),
we say vertex pair $(s,t)$ is covered by hub-set $\mathcal{H}$.
Let $C(\mathcal{H})$ be the number of (unique) vertex pairs covered by the hub-set $\mathcal{H}$. 
\edefin

Given this,  we are interested in maximally covering the shortest path distance using a limited number of vertices in the hub-set (in order to hold the distance matrix Hub$^2$). 
This can be formally described as follows: 

\bdefin{\bf (Optimal Hub-Set Selection Problem)}
Given graph $G=(V,E)$ and parameter $K$, we would like to find $\mathcal{H} \in V$,
such that $|\mathcal{H}|=K$ and the number of vertex pairs $(s,t) \in V \times V$ covered by $L$ is maximized ($\max{C(\mathcal{H})}$).
\edefin

Unfortunately, this problem is NP-hard. 
\bthm
The  optimal hub-set selection problem is NP-hard and it is also equivalent to the $k$-landmark-cover problem~\cite{Potamias09}. 
\ethm

The proof is in the appendix. 
Given this, we consider how to heuristically select vertices in the hub-set. It is quite similar to the landmark selection though here we focus on the coverage criterion on the social network. 
Let us start from the most simple scenario to gain a better intuition of good hub-set in social networks.
Intuitively, the best hub-set vertices should be the one in the very {\em central} position of graph since they tend to cover a large number of shortest paths~\cite{Potamias09}.
This high-level understanding suggests two concepts of {\em central} vertex from different perspectives:
1) it is a vertex lying on many shortest paths, which can be measured by vertex betweenness centrality;
2) it is a vertex close to other vertices in the graph, which can be measured by closeness centrality.
In the optimized scenario, the best {\em central} vertex based on first concept is referred to as the vertex with largest vertex betweenness~\cite{citeulike:1025135},
while second one refers to the vertex with optimal closeness centrality.
We generalize both concepts to hub-set selections in social networks.

\noindent{\bf Vertex Betweenness Criterion: }
The vertex betweenness firstly introduced by Freeman~\cite{citeulike:1025135} in the context of social science:
\[  C_B(v) = \sum_{u,w \in V \wedge u\neq w \neq v} \frac{\sigma_{uw}(v)}{\sigma_{uw}} \]
where $\sigma_{uw}(v)$ is the total number of shortest paths between $u$ and $w$ that pass through $v$,
and $\sigma_{uw}$ is the total number of shortest paths between $u$ and $w$.

We sort all vertices based on their vertex betweenness centrality in descending order and pick top $K$ vertices as hub-set.
To achieve this goal, we need to precompute vertex betweenness centrality for every vertex.
However, the fastest (exact) algorithm to compute every vertex betweenness centrality is $O(|V||E|)$~\cite{Brandes2001}, which is still rather expensive for massive networks. 

\noindent{\bf Closeness Betweenness Criterion: }
The closeness of a vertex is evaluated by the inverse of its total distance to all other vertices in the graph~\cite{citeulike:7062391}.
The vertex is the more central the lower its total distance to the remaining vertices. That is,
\[ C_C(v) = \frac{|V|-1}{ \sum_{u \in V \wedge u \neq v} d(v,u) }  \]

The intuition of using this criterion is that, a vertex closer to all other vertices are very likely to be part of many shortest paths.
Therefore, we select top $K$ vertices with largest closeness centrality to be hub-set.
Similar to computing vertex betweenness, calculating closeness betweenness involving pairwise shortest paths computation is an expensive task,
taking $O(|V|(|V|+|E|))$ time in unweighted graph setting.

\noindent{\bf Hub-Set=\{Hub\}:} 
Although it is costly to select $K$ hub-set vertices strictly following aforementioned criteria, 
we are able to adopt {\em the vertices with high vertex degree}, i.e., the {\em hubs}, as their effective proxies in social networks.
This is based on key observations on how high-degree vertices (hubs) contribute social networks' connectivity and small path length. 
The empirical study on several well-known social networks (i.e., Flickr, LiveJournal, Orkut and Youtube) reported in~\cite{Mislove07}
clearly demonstrates the presence of densely connected {\em core} comprising of a small portion of vertices with highest degree.
The removal of this {\em core} dramatically disconnects entire graph, implying that most of pairwise connections rely on few high-degree vertices (hubs).
In other words, few high degree vertices serve as key intermediates in most of shortest paths, thus are qualified to be the hub-set vertices under vertex betweenness criterion.
In addition, this study ~\cite{Mislove07} also analyzes the average shortest path length in the core containing only highest-degree vertices. 
As core grows, the average shortest path length increases sub-logarithmically.
Since average shortest path length of entire graph is very small, this phenomenon suggests that the lower degree vertices that constitute the majority of the network are very close to the high degree vertices in the core.
In this sense, high-degree vertices (hubs) are also qualified under the closeness betweenness criterion.

To sum, we simply select the top $K$ high-degree vertices (hubs) in the original graph $G$ as the hub-set ($\mathcal{H}$) (That is also the reason why we refer to $\mathcal{H}$ as the hub-set). 
In addition, we note that another advantage of using the hubs in the hub-set is to reduce the search space. 
Recall that in the bidirectional search (the query processing step), any vertices in the hub-set do not need to be expanded. 
In this way, we can effectively avoid those high-degree vertices (hubs), which form the very reason for search space explosion. 
Finally, in the rest of the paper, we will also use  the hub-set vertices and the hubs exchangeably.

\subsection{Core-Hub Labeling}
\label{indexingalgo}

In this subsection, we present an efficient algorithm to assign each vertex $u$ its core hubs $L_{H}(u)$.  
The goal of $6$-degree shortest path is to compute the shortest path with length no greater than $6$ (or $k$ in general).
Given this, it is easy to see that {\em for any hub $h$ in $\mathcal{H}$ to be assigned to $L_H(u)$, their distance has to be no greater than $6$ ($d(h,u) \leq 6$). }
This is because for any vertex pair $(s,t)$, the estimation using the hub-set is to provide an upper-bound  for the bidirectional search, which is also constrained by $6$. 
Thus, if the upper-bound is greater than $6$, it becomes useless.

Now, let us consider the straightforward approach based on the core-hub definition (Definition~\ref{corehubdef}) for labeling.
In this case, we need  first precompute all pairwise distances among hubs and materialize all distances between hubs and other vertices if they are no greater than $6$. 
Since the size of hub-set $\mathcal{H}$ is quite large, the materialization cost of $|V||\mathcal{H}|$ is too high. 

Here, we introduce an efficient procedure which only performs one local breadth-first traversal (BFS) for each hub in $\mathcal{H}$ to visit its $6$-degree neighbors(i.e., the vertices with distance to the hub no more than $6$) and no additional distance-matrix materialization is needed. 
In addition, this procedure also construct the distance matrix Hub$^2$ during the labeling process. 
Given this, for each BFS starting from a hub $h \in \mathcal{H}$, two tasks need to be performed: 
1) it needs to find out all other hubs whose distance to $u$ is no more than $6$, and record one shortest path between them (distance matrix construction);
2) for each visited non-hub vertex $v$ during BFS, we need to quickly check whether $h$ is $v$'s core hub, i.e., whether $u \in L_H(v)$.
This is equivalent to determine whether there is other hubs appearing in a shortest path between $u$ and $v$.
The first task can be easily achieved by introducing an extra integer for each vertex, which is used to record the previous vertex on the shortest path from hub $h$ to current vertex along shortest path.
The first task is relatively straightforward as a basic BFS process can easily accomplish that. 
The second task is non-trivial and is the main focus.

To quickly determine whether $h \in L_H(v)$ during the BFS process, we utilize the following observation:
{\em Let vertex $y$ lies on a shortest path from hub $h$ to non-hub vertex $v$ with distance $d(h,v)-1$ (i.e., $y$ is one hop closer than $v$ with respect to $h$). 
If there is a hub $h^\prime$ appearing in a shortest path from $h$ to $y$ ($h^\prime$ and $y$ may not be distinct),
$h^\prime$ definitely lies on a shortest path from $h$ to $v$ and $h$ cannot be core hub of $v$.}
Based on this observation, we simply maintain a binary flag $b(v)$ to denote whether there is another hub appearing in a shortest path between $h$ and $v$.
Specifically, its update rule is as follows: $b(v)=1$ if $v$ itself is hub or $b(y)=1$ where $d(h,y)=d(h,v)-1$ and $d(y,v)=1$.
Thus, during BFS traversal, when we visit vertex $v$, if its flag $b(v) \neq 1$ meaning there is no other hubs lying on the shortest path between $h$ and $v$, we add $h$ and the corresponding shortest path $SP(h,v)$ to its $L_{H}(v)$.

\begin{example}
Figure~\ref{fig:bfsindexexample} shows an example of core-hub labeling. 
The value attached to each vertex $v$ is its binary flag $b(v)$, such as $b(1)=0$ and $b(2)=1$.
As we know, BFS process visits vertices in levelwise manner with the help of queue. 
We start from hub $h$ to visit its neighbors $1$, $2$ and $3$ in level $1$ and push them ino queue.
When vertex $2$ is popped from queue, its binary flag $b(2)$ is set to be $1$ since it belongs to hub-set.
Afterward, the flags of its immediate neighbors in level $2$ (i.e., vertices $5$ and $7$) are all set to be $1$,
because their neighbor $2$ in level $1$ is hub and lies on shortest paths from hub $h$ to them.
Therefore, hub $h$ cannot be core-hub of $5$ and $7$.
However, the flag of vertex $6$ is $0$ since flags of all its immediate neighbors in level $1$ are $0$,
meaning no hub lying on shortest paths from $h$ to $6$,
thus $h \in L_H(6)$.
\end{example}

\begin{figure}
\centering
\begin{tabular}{c}
\psfig{figure=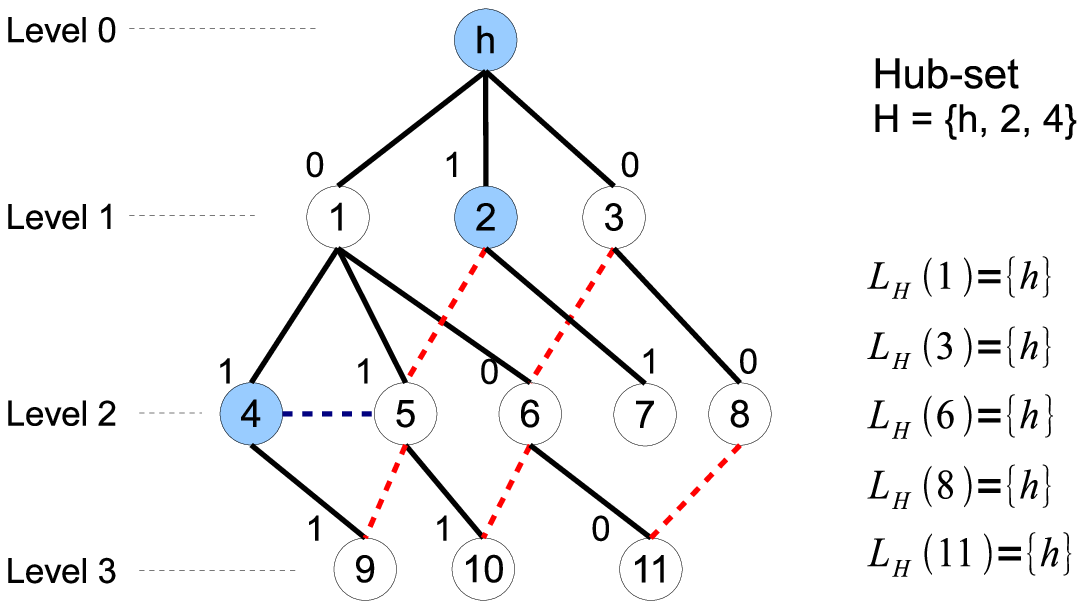,scale=0.65}
\end {tabular}
\caption{Example for Core-hub Labeling (only consider 3 hops)}
\label{fig:bfsindexexample}
\end{figure}

\begin{algorithm}
{\small
\caption{BFSIndexing($G=(V,E)$,$h$,$\mathcal{H}$)}
\label{alg:bfsindex}
\begin{algorithmic}[1]
\STATE Initialize $b(u)\leftarrow 0$ for each vertex $u$;
\STATE $level(h) \leftarrow 0$; $Q \leftarrow \{h\}$ \COMMENT{queue for BFS};
\WHILE{$Q \neq \emptyset$}
    \STATE $u \leftarrow Q.pop()$;
    \IF{$u  \in \mathcal{H}$ and $level(u) \ge 1$}
        \STATE $b(u) \leftarrow 1$;
        \STATE record shortest path $SP(h,u)$ from hub $h$ to hub $u$;
    \ENDIF
    \IF{$b(u)=0$ \COMMENT{no other hubs appear in any shortest path $SP(s,u)$}}
        \STATE $L_{H}(u) \leftarrow L_{H}(u) \cup \{<h:level(u),SP(h,u)>\}$;
    \ENDIF
    \FORALL{$v \in neighbor(u)$ \COMMENT{$(u,v) \in E$}}
        \IF{$level(u) < 6$ \COMMENT{$d(h,v)< 6$}}
            \IF{$v$ is not visited}
                \STATE $Q.push\_back(v)$;
            \ENDIF
            \IF{$level(v) = level(u)+1$ and $b(u)=1$}
                \STATE $b(v) \leftarrow 1$;
            \ENDIF
        \ENDIF
    \ENDFOR
\ENDWHILE
\end{algorithmic}
}
\end{algorithm}

The outline of BFS-based procedure for generating core-hubs indices is described in Algorithm~\ref{alg:bfsindex}.
The flag $b(v)$ for each vertex $v$ is updated when $v$ is hub or the flag of its previous vertex in $SP(h,v)$ is one (Line 6 and Line 18).
When vertex $u$ is popped from queue and its flag $b(u)=0$, hub $s$ is assigned to $L_{H}(u)$ (Line 10).
This procedure will be invoked for each hub.
In total, this procedure takes $O(\sum_{h \in \mathcal{H}} (|N_6(h)|+|E_6(h)|))$ time, where $L$ is hub-set, $N_6(h)$ ($E_6(h)$) is the number of vertices (edges) in $u$'s 6-degree neighborhood.
}



\comment{
\section{Local Landmark Framework}
\label{landmark}

In this section, we describe a new indexing scheme using landmarks to effectively encode the shortest path information.
Given a landmark set, new indexing scheme not only provides much compact indices, but also guarantees the resulting shortest path no long than the one by traditional landmark approach.
We formally define the optimal landmark selection problem based on new indexing scheme.
Then, we show that this problem is NP-hard and propose a greedy solution based on set-cover-with-pair framework to achieve logarithmic approximation.

To facilitate our discussion, we begin with defining some nation and terminology used in the rest of paper.
Let $G=(V,E)$ be an unweighted graph where $V=\{1,2,...n\}$ is the vertex set and $E \subseteq V \times V$ is edge set.
The edge from vertex $u$ and $v$ is denoted by $(u,v)$, and we use $P(v_0, v_p) = (v_0, v_1, ..., v_p)$ to denote a simple path from $v_0$ to $v_p$.
The length of a simple path in unweighted graph is the number of edges in the path.
Given two vertices $u$ and $v$, their shortest path $SP(u,v)$ is the path between them with minimal length.
The distance from vertex $u$ to $v$ is the length of shortest path $SP(u,v)$ from $u$ to $v$, denoted by $d(u,v)$.
For each vertex $u$, its $\epsilon$-neighbors, denoted as $N_{\epsilon}(u)$ is a set of vertices such that their distance to $u$ is no greater than $\epsilon$, i.e., $N_{\epsilon}(u) = \{v \in V | 0 < d(u,v) \le \epsilon\}$.
Given a vertex set $V^\ast$, for each each vertex $u \notin V^\ast$,
$Succ(u|V^\ast)$ is a subset of $V^\ast$ which can be reached from $u$ and $Pred(u|V^\ast)$ is a subset of $V^\ast$ which can reach $u$.
Due to the popularity of directed graphs in social networks, we focus on the undirected and unweighted graphs in the paper.
Taking a undirected graph as a bidirected graph, all techniques discussed in the paper can be easily extended to undirected setting.

\subsection{Global Landmark Indexing}
\label{globalindex}

Given graph $G=(V,E)$, suppose a subset of ordered vertices $V^\ast=\{x_1,x_2,...,x_k\} \subseteq V$ has been selected to be {\em landmarks}.
For each landmark $x_i$, we need to precompute and explicitly store the shortest path to all vertices in the graph $G$.
Then, we distribute these shortest path information to every vertex, such that each vertex $u$ is associated with two vectors:
{\small
\beqnarr
L_{in}(u)  = (<x_1: d(x_1,u),SP(x_1,u)>...,<x_p: d(x_p,u),SP(x_p,u)>)  \nonumber \\
L_{out}(u) = (<y_1: d(u,y_1),SP(u,y_1)>...,<y_q: d(u,y_q),SP(u,y_q)>) \nonumber
\eeqnarr
}
where $x_i \in Pred(u|V^\ast)$ ($1 \le i \le p$) and $y_j \in Succ(u|V^\ast)$ ($1 \le j \le q$).
Since all landmark information are propagated to entire graph, we refer to this traditional indexing scheme as {\em Global Landmark Indexing} scheme.
Given this, for a pair of vertices $s$ and $t$, their {\em landmark distance $d(s,t|V^\ast)$} and {\em landmark shortest path $SP(s,t|V^\ast)$} are defined based on triangle inequality:
{\small
\beqnarr
d(s,t|V^\ast) = min_{x \in L_{out}(s) \cap L_{in}(t)} \{ d(s,x)+d(x,t) \} \nonumber \\
SP(s,t|V^\ast) = SP(s,x^\ast) \circ SP(x^\ast,t) \nonumber \\
 \text{ where } x^\ast = \underset{x \in L_{out}(s) \cap L_{in}(t)}{\operatorname{argmin}} d(s,x)+d(x,t) \nonumber
\eeqnarr
}
where $\circ$ is the concatenation operator for two paths by connecting the first path's ending node and second path's starting node.
Clearly, given landmark set $V^\ast$, the landmark index size of total graph is $\sum_{u \in V} |L_{out}(u)|+|L_{in}(u)| = \sum_{u \in V} |Pred(u|V^\ast)| + |Succ(u|V^\ast)|$.

When landmark distance is equivalent to distance in the graph $G$, the landmark shortest path is actually the exact shortest path for a pair of vertices.
We say a set of vertices are {\em exact landmarks} if the landmark distance of any pair of vertices is exact distance in the graph $G$.
Simply speaking, for any two reachable vertices, there always exists one vertex from {\em exact landmarks} appearing in one of the shortest paths between them.
The problem of discovering exact landmarks with smallest cardinality has been proved to be NP-hard~\cite{Potamias:2009:FSP:1645953.1646063}.
Due to the hardness of selecting exact landmarks, most of existing works concentrate on developing different heuristics utilizing certain graph properties, such as vertex centrality, degree distribution or graph diameter.
Few of them attempt to improve the accuracy from indexing perspective.

\begin{figure}
\centering
\begin{tabular}{c}
\psfig{figure=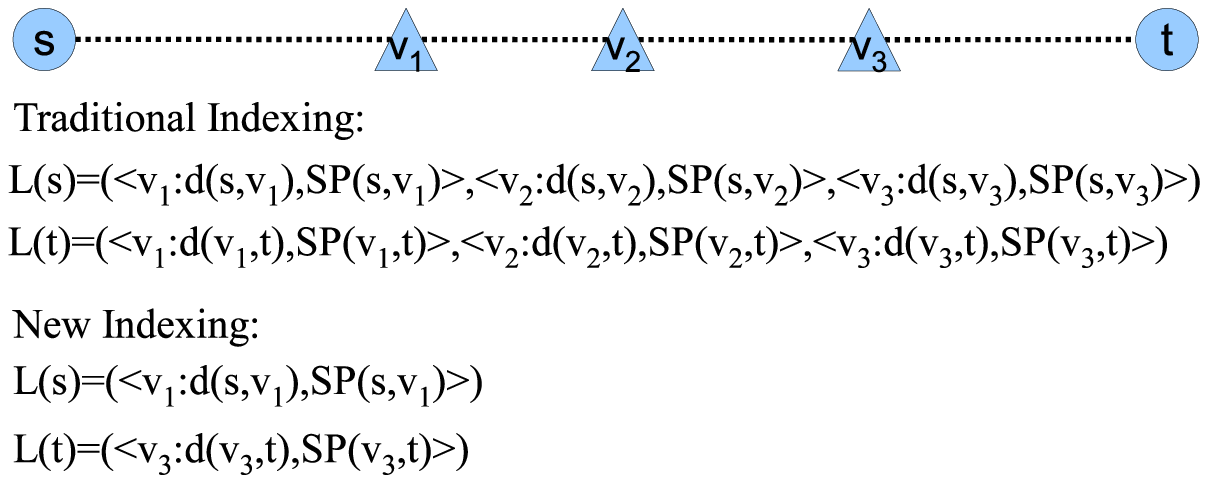,scale=0.6}
\end {tabular}
\caption {Indexing Example}
\label{fig:indexexample}
\end{figure}

\begin{figure}
\centering
\begin{tabular}{c}
\psfig{figure=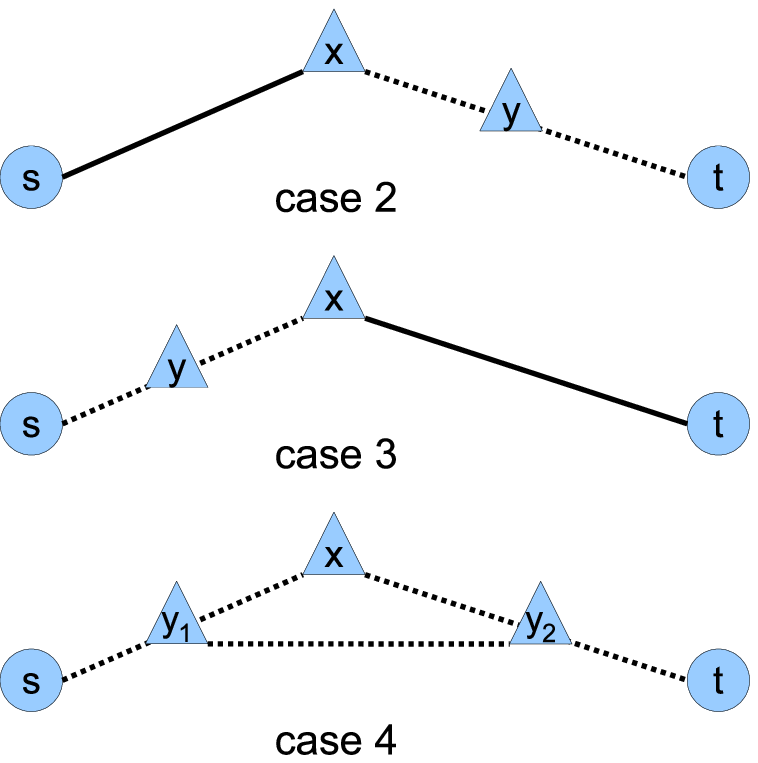,scale=0.65}
\end {tabular}
\caption {Cases in proof of Lemma~\ref{localgoodpath}}
\label{fig:proofcases}
\end{figure}

\subsection{Local Landmark Indexing}
\label{localindex}

In global landmark indexing scheme, each vertex is required to explicitly record all shortest distance and shortest path to (from) the landmarks which is reachable (can be reached).
However, we observe that many landmarks are redundant for each specific vertex under certain condition.
More specifically, each vertex only needs to record a subset of {\em essential} landmarks such that on other landmarks appear in the shortest path between them.
Using a slightly modified shortest path computation strategy, the recorded landmarks guarantee to generate a path which is no longer than the one produced by global indexing scheme.
Before we formally introduce new indexing scheme, let us go over a simple example to gain intuitive understanding.

\begin{example}
\label{indexexamp}
Figure~\ref{fig:indexexample} shows a shortest path from $s$ to $t$ while three landmarks $v_1$, $v_2$ and $v_3$ lie in this shortest path.
In global landmark indexing, both outgoing indices of source $s$ and incoming indices of destination $t$ need to record all three landmarks.
If shortest path $SP(v_1,v_3)$ between two landmarks $v_1$ and $v_3$ is known beforehand,
landmark $v_1$ included in $L_{out}(s)$ and landmark $v_3$ included in $L_{in}(t)$ are sufficient to recover the shortest path $SP(s,t)$,
i.e., $SP(s,t)= SP(s,v_1) \circ SP(v_1,v_3) \circ SP(v_3,t)$.
We can see that, the shortest path $SP(v_1,v_3)$ between two landmarks are implicitly recorded twice in $SP(s,v_3)$ and $SP(v_1,t)$,
thus materializing $SP(v_1,v_3)$ does not introduce any extra index cost.
\end{example}

In the example, we actually represent the shortest path $SP(s,t)$ by three segments:
1) $SP(s,v_1)$, from source $s$ to one close landmark $v_1$; 2) $SP(v_1,v_3)$, from landmark $v_1$ to another landmark $v_3$ close to destination $t$;
and 3) $SP(v_3,t)$, from landmark $v_3$ to destination $t$.
To some extent, two intermediate landmarks in the shortest path mimic the highway entrances or exits of transportation network.
In order to find the shortest path from starting point to destination, one simply find the shortest path from starting to the appropriate entrance of highway,
and then get off the correct exit of highway which leads to the remaining shortest path to the destination.
The shortest paths among landmarks serve as {\em shortest path oracle} which can answer the shortest path queries quickly, such as in constant time.

Now, given a landmark set $V^\ast$, we formally define {\em incoming essential landmarks} and {\em outgoing essential landmarks} for each vertex $u$:

\bdefin{\bf (Essential Landmarks)}
Given graph $G=(V,E)$ and a landmark set $V^\ast$, for each vertex $u$, its outgoing essential landmarks $L^e_{out}(u)$ contains landmark $x$,
if there does not exist $x^\prime \in V^\ast$ satisfying $d(u,x)=d(u,x^\prime)+d(x^\prime,x)$.
Simply speaking, no other landmarks appear in one of shortest path from $u$ to $x$.
Similarly, its incoming essential landmarks $L^e_{in}(u)$ contains landmark $y$,
if there is no $y^\prime \in V^\ast$ satisfying $d(y,u)=d(y,y^\prime)+d(y^\prime,u)$.
\edefin

The exclusion of a landmark $x$ in the essential landmarks with respect to vertex $u$ means there exists other landmark appear in the shortest path $SP(u,x)$ and closer to $u$ than $x$.
Recalling example~\ref{indexexamp}, $v_3$ is non-essential for source $s$, since the shortest path from $s$ to $v_3$ can be ``blocked'' by $v_1$.
The {\em incoming essential landmarks} and {\em outgoing essential landmarks} of each vertex constitute the indices of new landmark indexing scheme.
Opposing to global landmark indexing which needs to store all reachable (can be reached) landmarks, we refer to this indexing scheme as {\em local landmark indexing}.
Based on local indexing, we are able to compute its corresponding {\em landmark distance $d^e(s,t|V^\ast)$} and {\em landmark shortest path $SP^e(s,t|V^\ast)$} are defined as:
\beqnarr
d^e(s,t|V^\ast) = min_{x \in L^e_{out}(s) \wedge y \in L^e_{in}(t)} \{ d(s,x)+d(x,y)+d(y,t) \} \nonumber \\
SP^e(s,t|V^\ast) = SP(s,x^\ast) \circ SP(x^\ast,y^\ast) \circ SP(y^\ast,t) \nonumber \\
\text{ where } x^\ast = \underset{x \in L^e_{out}(s) \wedge y \in L^e_{in}(t)}{\operatorname{argmin}} d(s,x)+d(x,y)+d(x,t) \nonumber
\eeqnarr
In this sense, we need to materialize the pairwise shortest path among landmarks.
However, this materialization does not impose burden on the memory cost due to two reasons.
First, the length of shortest path in the social networks we focus on in this work is typically rather small~\cite{Mislove07}.
In our experiments, we observe that the length of most shortest paths between landmarks is no more than 3.
Second, lots of shortest paths between landmarks are implicitly encoded multiple times in global landmark indexing (e.g., Figure~\ref{fig:indexexample}, $SP(v_1,v_3)$ is contained in $SP(s,v_3)$ and $SP(v_1,t)$).

In addition, it is not hard to see that, local landmark indexing is a generalization of global landmark indexing.
In other words, if there is no shortest path from non-landmark vertex to landmark (or from landmark to non-landmark vertex) is ``blocked'' by some landmarks,
local landmark indexing degrades to global landmark indexing.
Hence, the index size of local landmark indexing $\sum_{u \in V} (|L^e_{out}(u)| + |L^e_{in}(u)|)$ is no larger that of global indexing.
More importantly, we have following observation:

\blemma
\label{localgoodpath}
The shortest path generated by local landmark indexing is no longer than the best one generated by global landmark indexing.
\elemma

\bproof
We want to prove that, for any path generated by global landmark indexing, we are able to find a path no longer that it based on local landmark indexing.
Without loss of generality, consider reachable vertex pair $(s,t)$, let us denote one path obtained by global landmark indexing as $SP(s,x) \circ SP(x,t)$,
where $x$ is the intermediate landmark in the path (i.e., $x \in L_{out}(s)$ and $x \in L_{in}(t)$.
Now, we try to find a path based on local landmark indexing which is not longer than $SP(s,x) \circ SP(x,t)$ in the following cases:

\noindent 1) $x \in L^e_{out}(s)$ and $x \in L^e_{in}(t)$. It is trivial to generate the same path $P(s,x) \circ P(x,t)$.

\noindent 2) $x \in L^e_{out}(s)$ and $x \notin L^e_{in}(t)$. According to the definition of {\em essential landmarks}, there is some landmark $y \in L^e_{in}(t)$ appear in the shortest path from $x$ to $t$, i.e., $d(x,t)=d(x,y)+d(y,t)$ (case $2$ shown in Figure~\ref{fig:proofcases}).
Since the shortest path $SP(x,y)$ from landmark $x$ to $y$ is available beforehand, we are able to generate the path $SP(s,x) \circ SP(x,y) \circ SP(y,t)$ with the same length as $SP(s,x) \circ SP(x,t)$.

\noindent 3) $x \notin L^e_{out}(s)$ and $x \in L^e_{in}(t)$. Follow the similar logic in case 2 (case 3 in Figure~\ref{fig:proofcases}). We guarantee to have $y \in L^e_{out}(s)$ appear in $SP(s,x)$. Thus, we can generate path $SP(s,y) \circ SP(y,x) \circ SP(x,t)$ which is no longer than $SP(s,x) \circ SP(x,t)$.

\noindent 4) $x \notin L^e_{out}(s)$ and $x \notin L^e_{in}(t)$. That is, there are some other landmarks $y_1 \in L^e_{out}(s)$ and $y_2 \in L^e_{in}(t)$ appear in the shortest paths $SP(s,x)$ and $SP(x,t)$, respectively (case 4 shown in Figure~\ref{fig:proofcases}).
Moreover the shortest path $SP(y_1,y_2)$ between landmarks $y_1$ and $y_2$ are known.
Due to triangle inequality $d(y_1,y_2) \le d(y_1,x)+d(x,y_2)$, we are able to generate the shorter path $SP(s,y_1) \circ SP(y_1,y_2) \circ SP(y_2,t)$
compared to $SP(s,x) \circ SP(x,t)$.

Putting those together, in any case, we can find a path based on local landmark indexing no worse than the one generated by global landmark indexing.
Therefore, the statement follows.
\eproof

Local landmark indexing scheme not only provides much compact index size, but also guarantees the resulting shortest path no long than the one by global landmark indexing scheme.
On the other hand, due to the properties of social networks, such as small diameter and massive size,
the desired landmarks should be able to provide exact shortest path using compact index size.
Motivated by this requirement, we formally define the concept of landmark coverage and following exact landmark discovery problem:

\bdefin{\bf (Landmark Coverage)}
Given graph $G=(V,E)$ and landmark set $V^\ast \subseteq V$, for a vertex pair $(s,t)$,
we say $V^\ast$ {\em covers} vertex pair $(s,t)$ if there exists two landmarks $x,y \in V^\ast$ appearing in the shortest path from $s$ to $t$.
Note that, two landmarks may not be distinct, i.e., $x$ may be the same with $y$.
\edefin

\bdefin{\bf (Exact Landmark Set Discovery Problem)}
Given an unweighted and directed graph $G=(V,E)$, the exact landmark set discovery problem tries to find a set of landmarks $V^\ast \in V$
which can cover any pair of reachable vertices and the local landmark index size is minimized.
\edefin

\comment{
\bdefin{\bf (Exact Landmark Set Discovery Problem)}
Given an unweighted and directed graph $G=(V,E)$, the exact landmark set discovery problem tries to find a set of landmarks $V^\ast \in V$,
such that for any pair of reachable vertices $(s,t)$, their corresponding local indices satisfy:
\[ d^e(s,t|V^\ast) = d(s,t) \]
where the local landmark index size is minimized.
\edefin
}

\subsection{Transform to Set-Cover-with-Pairs}
\label{scpsolution}

In this section, we transform the exact landmark set discovery problem to set-cover-with-pair instance.
To be self-contained, we first briefly introduce the set-cover-with-pair problem~\cite{HS05}.

\bdefin{\bf (Set-Cover-with-Pairs Problem ~\cite{HS05})}
Let $U$ be the ground set and let $S = \{1, \ldots, M\}$ be a set of objects, where each object $i\in S$ has a non-negative cost $w_i$.
For every $\{i,j\}\subseteq S$, let $\mathcal{C}(i, j)$ be the collection of elements in $U$ covered by the pair $\{i, j\}$.
The objective of the set cover with pairs (SCP) problem is to find a subset $S^\prime \subseteq S$ such that
$\mathcal{C}(S^\prime) =\bigcup_{\{i,j\}\subseteq S^\prime} \mathcal{C}(i, j) = U$ with a minimum covering cost $\sum_{i\in S} w_i$.
We refer to the special case in which each object has a unit weight ($w_i=1$) as the \textbf{cardinality SCP} problem.
\edefin

Now we transform our exact landmark set discovery problem to the Set-Cover-with-Pairs problem as follows.
Let the ground set $U=TC(G)$, i.e., it contains any vertex pair $(u,v)$ in $G$, such that $u \rightarrow v$.
To cover a pair $(u,v)$ means we can recover one of their shortest paths in the new landmark indexing scheme.

In our directed graph setting, we categorize objects in $Q$ into two types: starting objects $Q^s$ and ending objects $Q^e$ with respect to two endpoints of each shortest path.
Each starting object $O^s_{u,x_i}$ in $Q$ corresponds to a tuple $(u,(x_i: d_{x_i}, SP_{x_i}))$, where $u$ is a starting vertex and $x_i$ is a candidate landmark can be reached from $u$.
Note that, $d_{x_i}= d(u,x_i)$ and $SP_{x_i}=SP(u x_i)$ are shortest distance and one shortest path from $u$ to $x_i$, respectively.
Similarly, each ending object $O^e_{y_j,v}$ in $Q$ corresponds to a tuple $((y_j: d_{y_j}, SP_{y_j}),v)$, where $v$ is ending vertex and $y_j$ is a candidate landmark can reach $v$.
Here, $d_{y_j}= d(y_j,v)$ and $SP_{y_j}=SP(y_j, v)$ are shortest distance and one shortest path from $y_j$ to $v$, respectively.
Now, we define $\mathcal(O_{u,x_i}, O_{y_j,v})$ for each pair of objects $O_{u,x_i}$ and $O(y_j,v)$.
Given any four vertices $(u,x_i,y_j,v)$, if $d(u,x_i)+d(x_i,y_j)+d(y_j,v)=d(u,v)$, then vertex pair $(u,v)$ is covered by objects pair $\{O^s_{u,x_i}, O^e_{y_j,v}\}$.
That is, $\mathcal{C}(O^s_{u,x_i}, O^e_{y_j,v})$ contains vertex pair $(u,v)$.
Otherwise, $\mathcal{C}(O_{u,x}, O_{y,v})=\emptyset$.
Note that, aforementioned four vertices may not be distinct, i.e., $u$ may be the same with $x_i$, $x_i$ may be the same with $y_j$, etc.
It is not hard to see that each $\mathcal{C}(O_{u,x_i}, O_{y_j,v})$ at most contains one vertex pair.

Further, we assign a unit cost to each object, i.e., $w(O_{u,x})=1$.
Then, if we can find a set of objects $Q^\prime \subseteq Q$ such that $\mathcal{C}(Q^\prime)$ covers the ground set,
we are able to recover all shortest path for any pair of reachable vertices in the graph.
Therefore, the union of landmark indices $\cup_{u\in V} (L^e_{out}(u) \cup L^e_{in}(u))$ is the exact landmarks with smallest local landmark index size.
By applying the best available algorithm for SCP proposed in~\cite{HS05}, we obtains a $O(\sqrt{N\log N})$ approximation ratio to exact landmark set discovery problem,
where $N$ is the cardinality of the ground set $N=|U|$, for the cardinality SCP problem.

However, in our problem setting, the ground set is $U=TC(G)$ and $N=|TC(G)|$, which in the worst case is $\frac{n(n-1)}{2}$.
Thus, the approximation bound is $O(n\sqrt{\log n})$, which is meaningless.
Furthermore, without any prior knowledge, any vertex can be candidate of landmarks, in order to construct the set-cover-with-pair instance,
we need to precompute and explicitly record the distance and shortest path between any pair of vertices.
Both running time and memory cost are prohibitively expensive for most of graphs.
}

\comment{
For the traditional landmark indexing, each vertex $u$ in directed graph is assigned with two lists $L_{out}(u)=(<d(u,v_1),SP_{u,v_1}>,<d(u,v_2),SP_{u,v_2}>,...,<d(u,v_k),SP_{u,v_k}>)$ and
$L_{in}(u)=(<d(v_1,u),SP_{v_1,u}>,<d(v_2,u),SP_{v_2,u}>,...,<d(v_k,u),SP_{v_k,u}>)$.
Note that, if some landmarks are not reachable from (or cannot reach) specific vertex, the corresponding distance and shortest path is set to be null.

All landmark selection strategies are motivated by the important observation:
for one vertex pair $u$ and $v$, if there exists one of shortest path between them going through one of landmarks, this shortest path can be recovered by these landmarks.
The ideal landmark set, refer to as {\em landmark cover}, are able to recover exact shortest path for any pair of vertices in the graph.
Although it is computationally intractable, most of landmark selection heuristics are still proposed to target this goal by utilizing graph's underlying properties.
In this section, we focus on investigating the essence of landmark cover which is able to guide us to the practical landmark selection.
Especially, we first describe a new landmark indexing scheme resulting in much compact index size compared to traditional landmark indices.
We formally define the optimal landmark cover discovery problem based on this indexing scheme, and show that this problem is NP-hard.
To address it, we propose an effective algorithm based on set-cover-with-pair (SCP) framework to discover landmark cover with logarithmic bound.

\subsection{Notation}
\label{notation}

To facilitate our discussion, we begin with defining some nation and terminology used in the rest of paper.
Let $G=(V,E)$ be an unweighted and directed graph where $V=\{1,2,...n\}$ is the vertex set and $E \subseteq V \times V$ is edge set.
The edge from vertex $u$ and $v$ is denoted by $(u,v)$, and we use $P_{v_0, v_p} = (v_0, v_1, ..., v_p)$ to denote a simple path from $v_0$ to $v_p$.
The length of a simple path in unweighted graph is the number of edges in the path.
Given two vertices $u$ and $v$, their shortest path $SP_{u,v}$ is the path between them with minimal length.
The distance from vertex $u$ to $v$ is the length of shortest path $SP_{u,v}$ from $u$ to $v$, denoted by $d(u,v)$.
For each vertex $u$, its $\epsilon$-neighbors, denoted as $N_{\epsilon}(u)$ is a set of vertices such that their distance to $u$ is no greater than $\epsilon$, i.e., $N_{\epsilon}(u) = \{v \in V | 0 < d(u,v) \le \epsilon\}$.

Suppose a subset of vertices $V^\ast=\{v_1,v_2,...,v_k\} \subseteq V$ has been selected as landmarks.
For the traditional landmark indexing, each vertex $u$ in directed graph is assigned with two lists $L_{out}(u)=(<d(u,v_1),SP_{u,v_1}>,<d(u,v_2),SP_{u,v_2}>,...,<d(u,v_k),SP_{u,v_k}>)$ and
$L_{in}(u)=(<d(v_1,u),SP_{v_1,u}>,<d(v_2,u),SP_{v_2,u}>,...,<d(v_k,u),SP_{v_k,u}>)$.
Note that, if some landmarks are not reachable from (or cannot reach) specific vertex, the corresponding distance and shortest path is set to be null.

For any vertex pair $(s,t)$, the traditional landmark scheme approximate the shortest path by concatenating shortest paths $SP_{s,v^\prime}$ and $SP_{v^\prime,t}$ where $v^\prime \in V^\ast$ achieves minimal $d(s,v^\prime)+d(v^\prime,t)$.
Clearly, if $v^\prime$ lies in one of shortest path from $s$ to $t$, the estimated shortest path is exact one.
In ideal situation, for any vertex pair $(s,t)$, we can always find a landmark $x$ from landmark set $V^\ast$ such that $d(s,v)+d(v,t)=d(s,t)$.
We refer to such landmark set as {\em landmark cover}.
Furthermore, 2hop labels are essentially a {\em landmark cover},
because for any pair of vertices, there always exists one common vertex in their corresponding label such that at least one of their shortest paths goes through it.
These common vertices serves as {\em landmark cover} in our scenario.

\comment{
Here, we formally define the {\em landmark cover} and {\em optimal landmark cover discovery problem} based above indexing scheme as follows:

\bdefin{\bf (Landmark Cover)}
Given unweighted and directed graph $G=(V,E)$, and a subset of vertices $V^\ast \subseteq V$, we say $V^\ast$ is a landmark cover if and only for any vertex pair $(s,t)$,
there exists a landmark $v \in V^\ast$
\edefin

\bdefin{(\bf (Optimal Landmark Cover Discovery Problem)}
Given unweighted and directed graph $G=(V,E)$, we would like to seek a landmark cover with smallest number of vertices.
\edefin
}

\begin{figure}
\centering
\begin{tabular}{c}
\psfig{figure=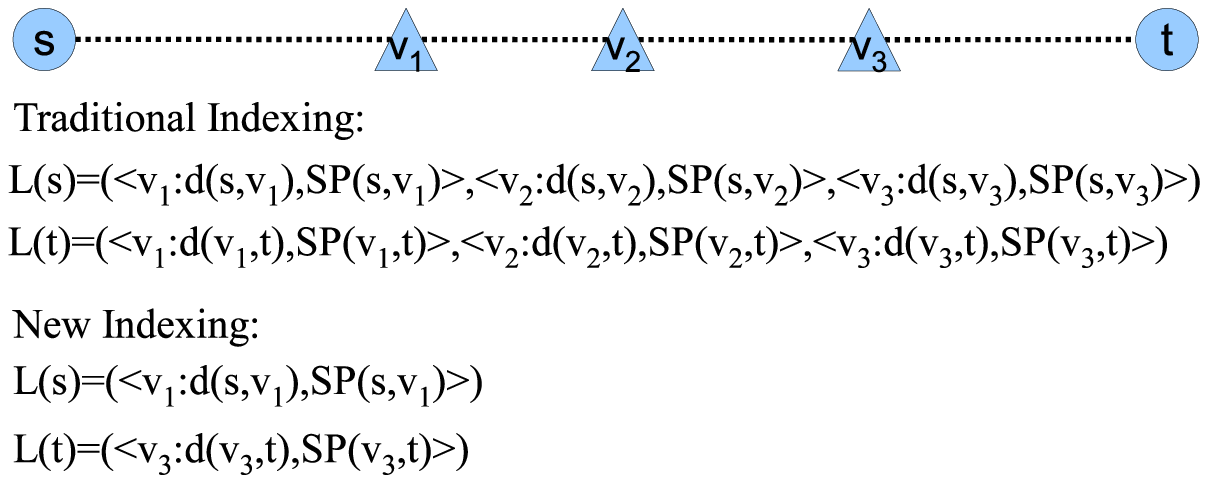,scale=0.65}
\end {tabular}
\caption {Indexing Example}
\label{fig:indexexample}
\end{figure}

\subsection{New Landmark Indexing Scheme}
\label{landmarkindex}

Considering a traditional landmark cover $V^\ast$ with $k$ landmarks, each vertex is required to explicitly maintain two $k$-element lists for the shortest path to (from) all landmarks.
Given a graph $G=(V,E)$, the total index size is clearly $2k|V|$.
However, we observe that many unnecessary landmarks are recorded for each specific vertex pair when one of their shortest paths goes through several landmarks.
In order to gain more compact index size,
we represent each shortest path as three segments instead of two segments in traditional methods: 1) the path from source to appropriate landmark close to it; 2) the path from current landmark to another landmark close to destination;
3) and the path between destination and above nearby landmarks.

We note that the functionality of these landmarks in this scenario is analogous to the highway entrances or exits of transportation network.
In order to find the shortest path from starting point to destination, one simply find the shortest path from starting to the appropriate entrance of highway,
and then get off the correct exit of highway which leads to the remaining shortest path to the destination.
Based on 3-segment expression, for any pair of vertices $(s,t)$, if there are many landmarks from one landmark cover appear in one of shortest path,
both $s$ and $t$ only need to store the {\em essential} landmark, i.e., the nearest one.
To facilitate the exact shortest path query, the middle segment connecting two landmarks are required to be available during online query.
As we discussed in section~\ref{contribution}, the landmark size is rather small and the shortest path among them is almost constant (no more than 3 in large social network),
thus the benefit from index size reduction would easily pay off the cost of materializing pairwise shortest path among landmarks.
This materialization essentially serves as {\em shortest path oracle} which is able to answer shortest path queries between any two landmarks in constant time.

\begin{example}
Figure~\ref{fig:indexexample} shows two different indexing schemes for vertex pair $(s,t)$ given that $3$ landmarks $v_1$, $v_2$ and $v_3$ appear in one of shortest path between them.
As we can see from traditional indexing, the shortest paths from $s$ to all landmarks and from landmarks to $t$ are all recorded.
According to 3-segment representation of shortest path, we only need to record one nearest landmark for each vertex, i.e., $v_1$ for source $s$ and $v_3$ for destination $t$.
The traditional landmark index size is reduced from $6$ to $2$.
Given the shortest path from $v_1$ to $v_3$, the shortest path $SP_{st}$ is easily recovered by $SP_{sv_1}+SP_{v_1 v_3}+SP_{v_3 t}$.
\end{example}

Our new landmark indexing scheme assigns each vertex $u$ two lists, $L^\prime_{out}(u)$ and $L^\prime_{in}(u)$.
$L^\prime_{out}(u)$ records the {\em essential} list of landmarks which is reachable from $u$, including their respective distance and shortest path,
i.e., $L^\prime_{out}(u)=(<x_1:d_{x_1},SP_{x_1}>,...,<x_p:d_{x_p},SP_{x_p}>)$, where $d_{x_i}=d(u,x_i)$ and $SP_{x_i} = SP_{u x_i}$.
Similarly, $L^\prime_{in}(u)$ records the {\em essential} list of landmarks which can reach $u$, including their respective distance and shortest path,
i.e., $L^\prime_{in}(u)=(<y_1:d_{y_1},SP_{y_1}>,...,<y_p:d_{y_p},SP_{y_q}>)$, where $d_{y_i}=d(y_i,u)$ and $SP_{y_i} = SP_{y_i u}$.
Both $x_i$ ($1 \le i \le p$) and $y_j$ ($1 \le j \le q$) are landmarks.
Given a shortest path query $(s,t)$, let $x_i \in L^\prime_{out}(s)$ and $y_i \in L^\prime_{in}(t)$.
We define the {\em landmark distance} from $s$ to $t$:
\beqnarr
d(s,t|L^\prime_{out}(s),L^\prime_{in}(t)) =  \nonumber \\
min_{x_i \in L^\prime_{out}(s), y_j \in L^\prime_{in}(t)} d(s,x_i)+d(x_i,y_j)+d(y_j,t) \nonumber
\eeqnarr
Corresponding {\em landmark shortest path} is the concatenation of $SP_{s x_i}$, $SP_{x_i y_j}$ and $SP_{y_j t}$.
Note that, the pairwise distance and shortest path among landmarks $d_(x_i y_j)$ and $SP_{x_i y_j}$ are materialized.
In addition, the index size is defined to be $\sum_{u \in V} (|L^\prime_{out}(u)| + |L^\prime_{in}(u)|)$.

If {\em landmark distance} by landmark set $V^\ast$ is exactly shortest distance for any pair of vertices,
we refer to $V^\ast$ as {\em landmark cover} in this setting.
Formally, we define {\em landmark cover} and {\em optimal landmark cover discovery problem} based new indexing scheme as follows:

\bdefin{\bf (Landmark Cover)}
Given unweighted and directed graph $G=(V,E)$, and a subset of vertices $V^\ast \subseteq V$, we say $V^\ast$ is a landmark cover if and only for any vertex pair $(s,t)$,
there exists two landmarks $x, y \in V^\ast$ such that $d(s,x)+d(x,y)+d(y,t)=d(s,t)$.
That is, there always exists two landmarks appearing in one of shortest paths from $s$ to $t$.
\edefin

When $x$ is the same with $y$, the landmark cover based on new indexing scheme is degraded to the one based on traditional indexing scheme,
which is also equivalent to the aforementioned 2-hop labels.

\bdefin{(\bf (Optimal Landmark Cover Discovery Problem)}
Given unweighted and directed graph $G=(V,E)$, we would like to seek a landmark cover such that its total index size based on new indexing scheme
$\sum_{u \in V} (|L^\prime_{out}(u)| + |L^\prime_{in}(u)|)$ is minimized.
\edefin

Since 2hop labeling problem can be treated as special case of Optimal Landmark Cover Discovery Problem, where two landmarks that serve as meeting point become only one.
we can easily derive following result:

\bthm
Given unweighted and directed graph $G=(V,E)$, Optimal Landmark Cover Discovery Problem is NP-hard problem.
\ethm

}

\vspace*{-2.0ex}
\section{Hub$^2$-Labeling for Shortest Path Computation}
\label{Hub2Labeling}

In this section, we present a Hub$^2$-labeling approach which aims to completely avoid visiting (and expanding) any hub. To achieve this, more expensive though often affordable pre-computation and memory cost are utilized for faster online querying processing. In Subsection~\ref{hub2labelingarchitect}, we will describe the Hub$^2$-labeling framework and its index construction.  In Subsection~\ref{HPBBFS}, we will discuss the faster bidirectional BFS.


\subsection{Hub$^2$-Labeling Framework}
\label{hub2labelingarchitect}
Hub$^2$-Labeling replaces the Hub-Network with a {\em Hub$^2$} distance matrix and {\em Hub Labeling}.   

\noindent{\bf Hub$^2$:}  The distance matrix between hub pairs  (referred to as   Hub$^2$) is precomputed and stored in main memory. Indeed, only the distances of pairs with distance no more than $k$ need to be computed for $k$-degree shortest path. As we discussed before, nowadays a desktop computer with moderate memory size can easily hold such a matrix for   $10K$ (or more) of hubs.  

\noindent{\bf Hub Labeling:} In order to effectively utilize the distance matrix, each vertex $v$ in the graph also records a small portion of hubs, referred to as the {\em core-hubs}, along with the distances. 
Basically, those core-hubs along with the distance matrix can help quickly estimate the upper-bound of distance between the query vertex pairs and can be used for bounding the search step of bidirectional BFS. 


Now, we formally define the {\em core-hubs}. 

\bdefin{\bf (Core-Hubs)}
\label{corehubdef}
Given graph $G=(V,E)$ and a collection $H$ of hubs, for each vertex $v$,
we say vertex $h \in H$ is a core-hub for $v$ 
if there is no other hub $h^\prime \in H$ such that $d(v,h)=d(v,h^\prime)+d(h^\prime,h)$.
Formally, $L(v)=\{h \in \mathcal{H}: \nexists h^\prime \in \mathcal{H}, d(v,h)=d(v,h^\prime)+d(h^\prime,h)\}$. 
\edefin

Simply speaking, if no other vertex $h^\prime$ appears in any shortest path between $v$ and $h$, $h$ is $v$'s core-hub. 
Note that a pair $(v,h)$, where $v \in L(v)$, is similar to a basic pair in the hub-network (Subsection~\ref{hubnetworkdiscovery}). The original basic pair definition only refers to hub pairs, but here it is being extended to vertex pairs with one hub and one non-hub vertex. 

\begin{example}
Figure~\ref{runningexample}(c) illustrate the core-hubs (along with the distance) for each non-hub vertices in the original graph (Figure~\ref{runningexample}(a)). 
Here the hubs are $4$, $6$, $8$, $12$, $17$, $18$, and $19$. 
For instance, Vertex $1$ only needs to record core-hubs $4$, $6$, $12$ and $19$, and it can reach hubs $8$ and $17$ through them in some shortest path. 
\end{example}

Using the core-hubs  $L$ and distance-matrix Hub$^2$ , we can approximate the distance and the shortest path for vertex pair $(s,t)$ in the following fashion: 
\beqnarr
d_H(s,t) = min_{x \in L(s) \wedge y \in L(t)} \{ d(s,x)+d(x,y)+d(y,t) \}  \label{corepath}
\eeqnarr
Here, $d(x,y)$ is the exact distance recorded in the distance-matrix Hub$^2$. 
 

The construction of the distance matrix Hub$^2$ and the  labeling of core-hubs are also rather straightforward. 
The BFS procedure in Algorithm~\ref{alg:bfsextraction} can be easily adopted: 1) each BFS performs $k$ steps and thus the distance matrix can be directly constructed; 2) when a vertex $v$ has flag $b=1$  (basic pair) from BFS traversal of $h$, we simply append $h$ to $L(v)$. 
Thus, the total computational complexity of the pre-computation is $O(\sum_{h \in H} (N_k(h)+E_k(h)))$ time, where $H$ is the hub set and $N_k(h)$ and $E_k(h)$ are the number of vertices and edges, respectively, in $u$'s k-degree neighborhood. 
We note that for directed graphs, we will compute both $L_{in}(v)$ and $L_{out}(v)$, one for incoming core-hubs $(h,v)$  and the other for outgoing core-hubs $(v,h)$. 
To construct such labels, we need perform both forward and backward BFS from each hub. 

The overall memory cost of Hub$^2$-Labeling is the sum of the cost of the distance matrix (Hub$^2$) together with the core-hub labeling for each vertex ($L(v)$): $\sum_{v \in V} O(|L(v)|) + O(|H|^2)$. 
This turns out to be rather affordable. 
In the experimental study, we found that for most of the real social networks, the core-hubs of each vertex $v$ is only a small portion of the total hubs (in most case, less than or close to $2\%$). 
Thus, the  Hub$^2$-Labeling can easily handle graphs with more than $10K$ hubs. 
Furthermore, since the second term (the size of the distance matrix) is stable, as the number of vertices increases in the original graph, the first term will scale linearly with respect to $|V|$. 

\comment{
\begin{example}
Now, let us take network Flicker we used in experiments as example to get in-depth understanding.
The network contains around $1.7$ millions vertices and $22.6$ millions edges and we select $10K$ highest degree vertices as hub-set.
Using hub-indexing scheme, each vertex only needs to record around $232$ core-hubs (i.e., around $2.3\%$ of entire hub-set), and total number of entries for Hub$^2$ is at most $100$ millions.
Distributing materialization cost of Hub$^2$ to each vertex, each vertex at most needs to record around $290$ ($=232+\frac{100M}{1.7M}$) entries, which is still significantly smaller than $10K$ in traditional landmark approach.
\end{example}}

\subsection{Hub$^2$-Labeling Query Processing} 
\label{HPBBFS}

To compute the $k$-degree shortest path between vertex pair $(s,t)$, the online query process in Hub$^2$-Labeling consists of two steps:

\noindent{\bf Step 1 (Distance Estimation): } Using the distance matrix Hub$^2$ and core-hubs labeling $L(s)$ and $L(t)$, the distance  $d_H(s,t)$ is estimated (Formula~\ref{corepath}). 

\noindent{\bf Step 2 (Hub-Pruning Bidirectional BFS (HP-BBFS)):} A bidirectional BFS from $s$ and $t$ is performed and the search step is constrained by the minimum between $k$ (for $k$-degree shortest path) and $d_H(s,t)$. 
In particular, none of the hubs need to be expanded during the bidirectional search. 
Mathematically, the Hub-Pruning Bidirectional BFS is equivalent to performing a typical Bidirectional BFS on the non-hub induced subgraph, $G[V \setminus H]$ of $G$.

\comment{
The fast hub-avoidance bidirectional BFS is sketched in Algorithm~\ref{alg:step2query}.
Note that {\em BackwardSearch} is essentially the same as the {\em ForwardSearch} and is omitted for simplicity.  
Initially, the $dist$ is determined to be the smallest one between the length of shortest path $d_H(s,t|\mathcal{H})$ and $6$ (or $k$ in general) (Line 3).
The main loop employes a forward search and a backward search in an alternating manner (Lines $4-12$).
In the BFS search procedure {\em ForwardSearch} (and {\em BackwardSearch}), 
each hub in $\mathcal{H}$ is excluded from the queue to avoid expanding their neighbors (Line 9).
Note that, the typical stop condition in the bidirectional search is that one direction visit a vertex already scanned in the other direction.
To fully utilize $d_H(s,t|L)$ as upper bound, we adopt a better criterion: the search terminates when the sum of distance labels of top elements in queues of the forward search and the backward search is no less than the length of shortest path seen so far (Line 10). We note that this criterion is firstly introduced in~\cite{GoldbergW05}.
}

\bthm
The two-step Hub$^2$-Labeling query process can correctly compute the $k$-degree shortest path in graph $G$. 
\ethm
\bproof
We observe that any vertex pair with distance no more than $k$ can be categorized as: 
1) vertex pairs having at least one shortest path passing through at least one hub in $H$;
and 2) vertex pairs whose shortest paths never pass through any hub. 

For any vertex pair $(s,t)$ with distance no greater than $k$ ($d(s,t) \le k$), 
if there exists one hub $x^\prime \in \mathcal{H}$ satisfying $d(s,t)=d(s,x^\prime)+d(x^\prime,t)$,
then, we can always find $x \in L_{H}(s)$ and $y \in L_{H}(t)$ such that $d(s,t)=d(s,x)+d(x,y)+d(y,t)$.
In other words, Step $1$ (distance estimation), which uses the distance-matrix Hub$^2$ and core-hub labeling, can handle this category. 
Also, the Step $2$ will help confirm the shortest path belongs to this category (cannot find a shorter one). 

If an approximate shortest path computed in Step $1$ is not an exact one, then the shortest path does not involve any hub. Thus Step $2$ can guarantee to extract an exact shortest path using the bidirectional search in the non-hub induced subgraph $G[V\setminus H]$.
\eproof

The time complexity of online query processing of a pair $s$ and $t$ can be written as $O(|L(s)||L(t)|$ + $N_{k/2}(s|G[V\setminus H])$+ $E_{k/2}(s|G[V\setminus H])$+$N_{k/2}^\prime(t|G[V\setminus H])$+ $E_{k/2}^\prime(t|G[V\setminus H])$), where $|L(s)||L(t)|$ is the distance estimation cost and the remaining terms are the cost of bidirectional search. 
$N_{k/2}$ ($N^\prime_{k/2}$) and $E_{k/2}$ ($E^\prime_{k/2}$) are the number of vertices and edges in the $k/2$-neighborhood (reversed neighborhood which follows the incoming edges) of the non-hub induced subgraph $G[V\setminus H]$. Since the hubs are excluded, the cost of hub-pruning bidirectional BFS is significantly smaller than that on the original graph. 

However, if the number of core-labels is large, then the distance estimation can be expensive (a pairwise join on $L(s)$ and $L(t)$ is performed). To address this issue, the core-hubs in $L(u)$ can be organized in a level-wise fashion, each level corresponding to their distance to $u$, such as $L^1(u), L^2(u), \cdots L^k(u)$. Using such a level-wise organization, we can perform a much more efficient distance estimation: the pairwise joins first performed between $L^1(s)$ and $L^1(t)$; then on $(L^1(s),L^2(t))$, $(L^2(s),L^1(t))$, $(L^2(s),L^2(t))$, etc. 
Given this, let us denote $d$ to be the shortest path length obtained by pairwise join so far.
Assuming we are currently working on  $(L^p(s),L^q(t))$,
if $d < p+q$, then we terminate the pairwise join immediately. 
This is because it is impossible for $(L^{p^\prime}(s),L^{q^\prime}(t))$ to produce better results since $p^\prime+q^\prime \ge p+q > d$.
This early termination strategy based on the level-wise organization can help us effectively prune unnecessary pairwise join operations and improve the query efficiency.

\comment{
\begin{figure}
\centering
\begin{tabular}{c}
\psfig{figure=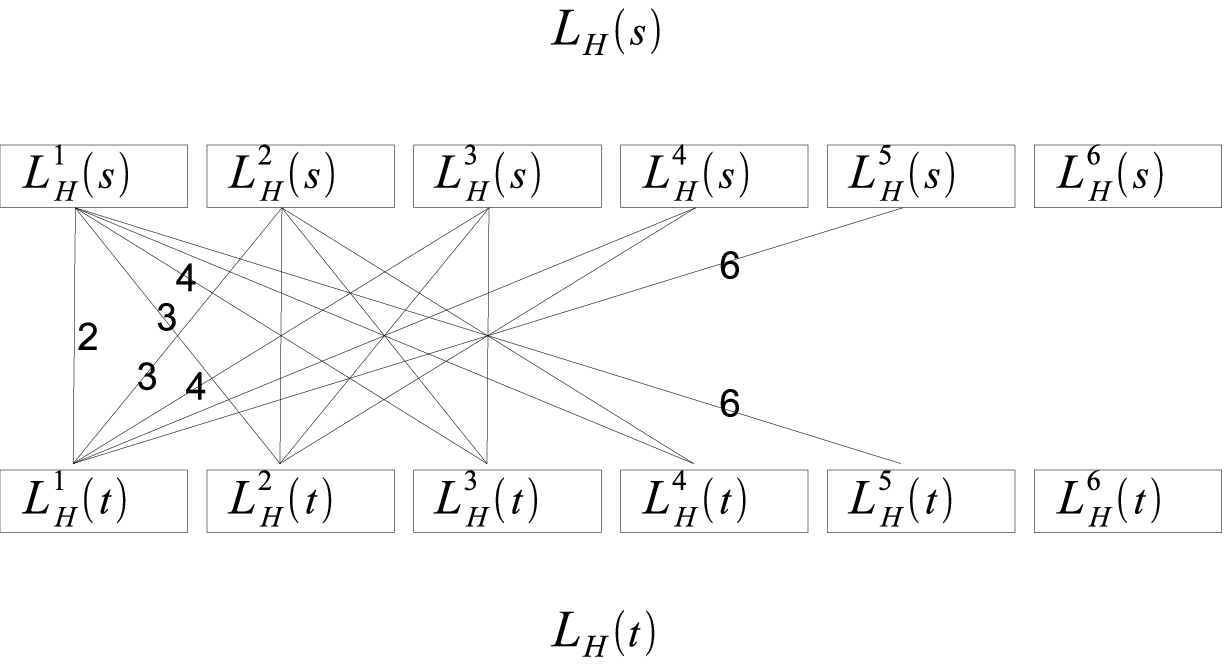,scale=0.65}
\end {tabular}
\caption{Levelwise Organization of Core-Hubs}
\label{fig:jointorder}
\end{figure}

\noindent{\bf Core-Hubs Levelwise Organization and Fast Distance Estimation:}
Given any vertex pair $(s,t)$ and their landmark indices $L_{H}(s)$ and $L_{H}(t)$,
the straightforward approach to estimate their shortest path is to perform a pairwise join on $L_{H}(s)$ and $L_{H}(t)$ based on Eq.~\ref{corepath}. This takes a total of $|L_{H}(s)| \times |L_{H}(t)|$  comparisons, which can be rather expensive.
To address this issue, we rank and store the core-hubs in $L_H(u)$ according to their distance to $u$ and utilize such structure for effectively pruning. 
\comment{
We observe that, when we check landmark pair $x \in L_{H}(s)$ and $y \in L_{H}(t)$,
if the shortest distance so far $dist$ is smaller than $d(s,x)+d(y,t)$, then we do not need to actually perform join comparison on $(x,y)$.
Further, any landmark pair satisfying such condition is not necessary to be accessed in the future, which potentially avoids many pairwise join operations.}

Specifically, we decompose  $L_{H}(u)$ into $6$ sublists $L^k_{H}(u)$,
where each sublist $L^k_{H}(u)$ contains hub $x_i$ with distance to $u$ is $k$, i.e., $d(u,x_i) = k$ ($1 \le k \le 6$).
Given this, pairwise join on $L_{H}(s)$ and $L_{H}(t)$ becomes pairwise join on two sets of sublists $L^k_{H}(s)$ ($1 \le k \le 6$) and $L^{k^\prime}_{H}(t)$ ($1 \le k^\prime \le 6$).
As shown in figure~\ref{fig:jointorder}, the connection between upper $L^p_{H}(s)$ and lower $L^q_{H}(t)$ indicates possible pairwise join on two sublists.
The value attached to each connection is the sum of distances $d(s,x)+d(y,t)=p+q$, where $x \in L^p_{H}(s)$ and $y \in L^q_{H}(t)$.
There is no connection between $L^4_{H}(s)$ and $L^4_{H}(t)$, because $d(s,x^\prime)+d(y^\prime,t)=8 > 6$ (i.e., the upper bound is $6$ in the $6$-degree shortest path query), where $x^\prime \in L^4_{H}(s)$ and $y^\prime \in L^4_{H}(t)$.

Using such a levelwise organization, we perform pairwise join on two sublists following the ascending order of value attached to corresponding connections between these two sublists (shown in Figure~\ref{fig:jointorder}).
Basically,  we first perform pairwise join on $L^1_{H}(s)$ and $L^1_{H}(t)$; then on $(L^1_{H}(s),L^2_{H}(t))$, $(L^2_{H}(s),L^1_{H}(t))$,$(L^2_{H}(s),L^2_{H}(t))$,..., etc. 
Given this, let us denote $d$ to be the shortest path length obtained by pairwise join so far.
Assuming we are currently working on landmarks from sublist pair $(L^p_{H}(s),L^q_{H}(t))$,
if $d < p+q$, then we terminate the pairwise join immediately. 
This is because the following sublist pair $(L^{p^\prime}_{H}(s),L^{q^\prime}_{H}(t))$ is impossible to produce better result since they have $p^\prime+q^\prime \ge p+q > d$.
This early termination strategy based on the levelwise organization can help us effectively prune unnecessary pairwise join operations and improve the query efficiency.
}

\comment{
Given landmarks $L$, for any vertex pair $(s,t)$, we are able to approximate their shortest path and distance by following expression:
{\small
\beqnarr
d^c(s,t|L) = min_{x \in L^c_{out}(s) \wedge y \in L^c_{in}(t)} \{ d(s,x)+d(x,y)+d(y,t) \} \nonumber \\
SP^c(s,t|L) = SP(s,x^\ast) \circ SP(x^\ast,y^\ast) \circ SP(y^\ast,t) \nonumber \\
\text{ where } x^\ast = \underset{x \in L^c_{out}(s) \wedge y \in L^c_{in}(t)}{\operatorname{argmin}} d(s,x)+d(x,y)+d(x,t) \nonumber
\eeqnarr
}
Note that, since this is only used to estimate the shortest path,
we do not consider the length constraint here, such that estimated path length is not greater than $6$.

Recall that our goal is to generate the {\em exact} shortest path with length no greater than $6$.
In the following, we introduce an important observation which would suggest...

Given the local landmark indices, we are only able to approximate the shortest path between any two vertices.
As any small error arisen by approximation is not acceptable in social networks with small diameters,
we aim at finding one exact shortest path for any vertex pair here.
To achieve this goal, we describe an efficient query processing procedure by leveraging local landmark indexing and bidirectional BFS.
Specifically, given vertex pair $(s,t)$, the query procedure typically involves two steps:

\noindent 1) we generate one approximated shortest path $SP(s,t|V^\ast)$ (distance $d(s,t|V^\ast)$) utilizing local landmark indices $L^e_{out}(s)$ and $L^e_{in}(t)$;

\noindent 2) a variant of bidirectional BFS is employed to complement inaccurate result generated by landmark-based method in order to obtain exact shortest path.

\begin{algorithm}
\caption{Query($s$,$t$)}
\label{alg:query}
\begin{small}
\begin{algorithmic}[1]
\REQUIRE{$L^e_{out}(s)$ and $L^e_{in}(t)$ are outgoing landmark indices and incoming landmark indices for $s$ and $t$} \\
\COMMENT{Step 1: local landmark-based approximation}
\STATE $dist \leftarrow \infty$; $P \leftarrow \emptyset$; \COMMENT{$P$ is the shortest path so far}
\FORALL{$x \in L^e_{out}(s)$}
    \FORALL{$y \in L^e_{in}(t)$}
        \IF{$d(s,x)+d(x,y)+d(y,t) < dist$}
            \STATE $dist \leftarrow d(s,x)+d(x,y)+d(y,t)$;
            \STATE $P \leftarrow SP(s,x) \circ SP(x,y) \circ SP(y,t)$;
        \ENDIF
    \ENDFOR
\ENDFOR \\

\COMMENT{Step 2: modified bidirectional BFS}
\STATE $forward \leftarrow true$; \COMMENT{flag used to indicate the search direction}
\STATE $Q_f \leftarrow \{s\}$; $Q_b \leftarrow \{t\}$; \COMMENT{$Q_f$ and $Q_b$ are queues for forward and backward search}
\WHILE{$Q_f \neq \emptyset$ or $Q_b \neq \emptyset$}
    \IF{$forward = true$}
        \STATE $(d_f, P_f) \leftarrow ForwardSearch(Q_f)$; \COMMENT{$d_f$ and $P_f$ are best distance and shortest path obtained from current forward search}
        \IF{$d_f < dist$}
            \STATE $dist \leftarrow d_f$;
            \STATE $P \leftarrow P_f$;
        \ENDIF
    \ELSE
        \STATE $(d_b, P_b) \leftarrow BackwardSearch(Q_b)$;
        \IF{$d_b < dist$}
            \STATE $dist \leftarrow d_b$;
            \STATE $P \leftarrow P_b$;
        \ENDIF
    \ENDIF
    \IF{$d(s,Q_f.top) + d(Q_b.top,t) \ge dist$ \COMMENT{$Q_f.top$ and $Q_b.top$ are top elements in both queues}}
        \STATE BREAK;
    \ENDIF
\ENDWHILE
\end{algorithmic}
\end{small}
\end{algorithm}

\comment{
\STATE $Q_f \leftarrow \{s\}$; $Q_b \leftarrow \{t\}$;
\COMMENT{$Q_f$ and $Q_b$ are queues for forward search and backward search, respectively}
\STATE $d \leftarrow \infty$;\\
\COMMENT{perform forward search and backward search}
\WHILE{$Q_f \neq \emptyset$ or $Q_b \neq \emptyset$}
    \STATE $(r,mp) \leftarrow ForwardSearch(Q_f)$;
    \IF{$r < d$}
        \STATE $d \leftarrow r$;
    \ENDIF
    \STATE $(r,mp) \leftarrow BackwardSearch(Q_b)$;
    \IF{$r < d$}
        \STATE $d \leftarrow r$;
    \ENDIF
\ENDWHILE \\
\COMMENT{backtrace shortest path}
\STATE $SP \leftarrow \emptyset$;
\STATE $u \leftarrow mp$;
\WHILE{$Prev_f[u] \neq \emptyset$}
    \STATE $SP.push\_back(u)$;
    \STATE $u \leftarrow Prev_f[u]$;
\ENDWHILE
\STATE $u \leftarrow Prev_b[mp]$;
\WHILE{$Prev_b[u] \neq \emptyset$}
    \STATE $SP.push\_front(u)$;
    \STATE $u \leftarrow Prev_b[u]$;
\ENDWHILE
\RETURN $(SP,d)$;

\begin{algorithmic}[1]
\PROCEDURE{ForwardSearch($Q_f$)}
\STATE $Q \leftarrow Q_f$; $Q_f \leftarrow \emptyset$; $r \leftarrow \infty$;
\STATE $mp \leftarrow -1$; \COMMENT{$mp$ is the meeting point between forward search and backward search}
\WHILE{$Q \neq \emptyset$}
    \STATE $u \leftarrow Q.pop()$;
    \FORALL{$v \leftarrow Neighbor(u)$}
        \IF{$v$ is visited in backward search}
            \IF{$l_f(u)+l_b(v)+1 < r$}
                \STATE $r \leftarrow l_f(v)+l_b(v)$;
                \STATE $mp \leftarrow v$;
            \ENDIF
        \ENDIF
        \IF{$v$ is not visited and $l_f(v) \le 3$}
            \IF{$v$ is gate vertex}
                \STATE $Q_f.push\_back(v)$;
            \ELSIF{$l_f(v)<3$}
                \STATE $Q.push\_back(v)$;
            \ENDIF
            \STATE $Prev_f[v] \leftarrow u$;
        \ENDIF
    \ENDFOR
\ENDWHILE
\RETURN $(r,mp)$;
\end{algorithmic}

}

Rather than traditional bidirectional BFS accessing the vertices from both direction without any constraint,
our new method utilizes the distance obtained from step 1 as upper bound to reduce the search space.
On the other hand, as we know, if the exact shortest path goes though some landmarks, it would be guaranteed to be found in the step 1.
Therefore, during BFS traversal procedure, we do not need to access landmarks which are essentially high-degree vertices in the social network.
This would further dramatically reduce the search space which is caused by accessing high-degree vertices with even hundreds of thousands of neighbors in the social network.
}

\vspace*{-2.0ex}
\section{Experimental Evaluation}
\label{exp}

In this section, we empirically evaluate the performance of our algorithm on a range of large real social networks.
In particular, we will compare the Hub-Network approach (denoted as {\bf HN}) and Hub$^2$-Labeling approach (denoted as {\bf HL}) with the following methods: 1) basic breadth-first search (denoted as {\bf BFS}); 2) bidirectional breadth-first search (denoted as {\bf BiBFS}); 3) the Sketch algorithm~\cite{SarmaSMR10} (denoted as {\bf S$^\star$}), the state-of-the-art approximate distance estimation algorithm; 4) the TreeSketch method~\cite{Gubichev:2010:FAE:1871437.1871503} (denoted as {\bf TS$^\star$}), which utilizes a tree to improve the  approximation accuracy of Sketch based shortest path computation. Here the symbol $\star$ also indicates it is an approximation method. 

In addition, we have also tested the two latest exact shortest path distance methods, including tree decomposition based shortest path computation~\cite{DBLP:conf/sigmod/Wei10} and the highway-centric labeling approach ~\cite{DBLP:conf/sigmod/JinRXL12} based on authors' provided implementation. However, neither of them can work on the graphs used in this study. This is as expected as their indexing cost is very high (tree decomposition or set-cover approach) and they are mainly focusing on very sparse graphs. 

We also tested RigelPath, another recent approach on approximate shortest path discovery in social networks~\cite{zhao2011}.
However, its query performance is slower than that of Sketch (also confirmed in their own study ~\cite{zhao2011}). 
Furthermore, its current implementation only focuses on undirected graphs, wheres most of the real benchmarking networks are directed. Thus, we do not report RigelPath's experimental results here.

\comment{
Note that, we do not compare our algorithm against rigel approach~\cite{Zhao11} due to following reasons.
First of all, rigel only focuses on undirected graphs while most of our real networks are treated to be directed.
More importantly, although it is very fast to estimate distance,
we found that the performance of estimating shortest path using rigel is much slower than Sketch on all datasets, as shown in their paper's Table VI.
Therefore, we adopt sketch-based approach as representative of state-of-the-art approximation approaches in our empirical study.}

We implemented our algorithms in C++ and the Standard Template Library (STL).
The implementation of  sketch-based approaches (including {\bf S$^\ast$} and {\bf TS$^\ast$}) is kindly provided by authors~\cite{Gubichev:2010:FAE:1871437.1871503} (also implemented in C++).
All experiments were run on a Linux server with 2.48GHz AMD Opteron processors and 32GB RAM.

In experiments, we are interested in two important measures: query time and preprocessing cost, which consists of precomputation time and indexing size.
To measure the query time, we randomly generate $10,000$ vertex pairs and obtain the average running time for each query.
For the index size, since all Sketch indices are stored in RDF format, their indexing sizes are measured in terms of the corresponding RDF file size.
If the preprocessing cannot be finished in $48$ hours,we will stop it and record ``-'' in the table of results.
Furthermore, we note that all Sketch-based benchmarks can only approximate shortest paths, where approximation accuracy is influenced by an iterative sampling procedure.
A parameter $r$ is specified to determine the number of sampling iterations, which leads to $2r\log |V|$ sketches for each vertex.
To make a fair comparison with exact query schemes, we set $r=2$ as suggested in~\cite{Gubichev:2010:FAE:1871437.1871503} which can produce sketches with good approximation accuracy and efficient query processing.
Also, in this study, we focus on comparing their query time again the new approaches despite they are only able to provide approximate solution whereas our approaches can provide the exact solution.

\begin{table*}[!ht]
\begin{minipage}[b]{0.38\linewidth}
\begin{center}
{\small
\begin{tabular}{|l|r|r|r|r|}
\hline
Dataset & \multicolumn{1}{c|}{$|V|$} & \multicolumn{1}{c|}{$|E|$} & \multicolumn{1}{c|}{$\overline{\delta}$} & \multicolumn{1}{c|}{$d_{0.9}$} \\ \hline


Facebook         &  63731     &  1545686     &   48.51    &   8.2   \\ \hline
Slashdot         &  82168     &  948464      &   23.09    &   4.7   \\ \hline
BerkStan         &  685230    &  7600595     &   22.18    &   10    \\ \hline
Youtube          &  1138499   &  4945382     &   8.69     &   7.14  \\ \hline
As-skitter       &  1696415   &  11095298    &   13.08    &   5.9   \\ \hline
Flickr           &  1715255   &  22613981    &   26.37    &   7.32  \\ \hline
Flickr-growth    &  2302925   &  33140018    &   28.78    &   7.19  \\ \hline
Wiki-talk        &  2394385   &  5021410     &   4.19     &   4     \\ \hline
Orkut            &  3072441   &  223534301   &   145.51   &   5.7   \\ \hline
LiveJournal      &  5204176   &  77402652    &   29.75    &   8.34  \\ \hline
Twitter          &  11316811  &  85331845    &   15.08    &  24.97  \\ \hline

\end{tabular}
}
\end{center}
\vspace*{-2.0ex}
\caption{Network Statistics}
\label{realdata}
\vspace*{-1.0ex}

\end{minipage}
\hspace{0.5cm}
\begin{minipage}[b]{0.52\linewidth}
{\small
\begin{tabular}{|l|r|r|r|r|r|r|r|r|}
\hline
\multirow{2}{*}{Dataset} & \multicolumn{2}{c|}{$|H|=5000$} & \multicolumn{2}{c|}{$|H|=8000$} &
\multicolumn{2}{c|}{$|H|=10000$} & \multicolumn{2}{c|}{$|H|=15000$} \\
\cline{2-9}
& HN & HL & HN & HL & HN & HL & HN & HL
\\ \hline
{ Facebook} & 0.043 & 0.018 & 0.044 & 0.017 & 0.042 & 0.017 & 0.040 & 0.019\\ \hline
{ Slashdot} & 0.023 & 0.002 & 0.021 & 0.001 & 0.022 & 0.001 & 0.022 & 0.002\\ \hline
{ BerkSta} & 0.011 & 0.005 & 0.005 & 0.009 & 0.010 & 0.004 & 0.014 & 0.002\\ \hline
{ Youtube} & 0.106 & 0.006 & 0.119 & 0.005 & 0.125 & 0.005 & 0.136 & 0.005\\\hline
{ As-skitter} & 0.051 & 0.016 & 0.044 & 0.015 &  0.040 & 0.013 & 0.041 & 0.011 \\ \hline
{ Flickr} & 1.600 & 0.112 & 1.671 & 0.073 & 1.739 & 0.067 & 1.888 & 0.061\\ \hline
{ Flickr-growth} & 0.998 & 0.138 & 1.130 & 0.113 & 1.193 & 0.100 & 1.236 & 0.136\\\hline
{ Wiki-talk} & 0.014 & 0.002 & 0.016 & 0.002 & 0.014 & 0.002 & 0.014 & 0.001\\ \hline
{ Orkut} & 0.952 & 3.653 & 0.955 & 3.314 & 0.978 & 3.356 & 1.078 & 3.282\\ \hline
{ LiveJournal} & 0.466 & - & 0.526 & - & 0.513 & - & 0.577 & -\\\hline
{ Twitter} & 1.850 & 0.306 & 1.947 & 0.314 & 2.083 & 0.340 & 2.121 & -\\ \hline

\end{tabular}
}
\vspace{-2.0ex}
\caption{{\small Average Query Time with Different Hub Sizes (ms)}}
\label{tab:hubsize}
\vspace*{-1.0ex}

\end{minipage}
\end{table*}

\begin{table*}[!ht]
\begin{minipage}[b]{0.45\linewidth}
{\small
\begin{tabular}{|l|r|r|r|r|r|r|}
\hline
\multirow{2}{*}{Dataset} & BFS & S$^*$ & TS$^*$ & BiBFS & HN & HL\\
\cline{2-7}
& \multicolumn{3}{c|}{$ms$} & \multicolumn{3}{c|}{$\mu s$}
\\\hline
Facebook & 1.7 & 0.5 & 20.4 & 55.2 & 41.9 & 17.4\\\hline
Slashdot & 1.4 & 0.7 & 34.5 & 31.6 & 22.2 & 1.3\\\hline
BerkStan & 0.3 & 4.7 & 559.1 & 33.9 & 10.2 & 3.5\\\hline
Youtube & 15.3 & 2 & 171.2 & 312.2 & 125.1 & 5.4\\\hline
As-skitter & 4.9 & 1.5 & 114.9 & 86.7 & 40.4 & 12.7\\\hline
Flickr & 42.6 & 2.7 & 288.7 & 2887.9 & 1738.8 & 67.3\\\hline
Flickr-growth & 71.8 & 5.1 & 305 & 1607.4 & 1193.3 & 100.3\\\hline
Wiki-talk & 18.8 & - & - & 56.4 & 14.1 & 1.5\\\hline
Orkut & 202.5 & 7.8 & 258.5 & 1338.7 & 978.1 & 3356.4\\\hline
LiveJournal & 131.4 & - & - & 749.6 & 513.1 & -\\\hline
Twitter & 221.4 & - & - & 2311.8 & 2082.6 & 339.7\\\hline
\end{tabular}
}
\vspace{-2.0ex}
\caption{{\small Average Query Time on Random Query}}
\label{tab:randquerytime}
\vspace*{-1.0ex}

\end{minipage}
\hspace{0.7cm}
\begin{minipage}[b]{0.45\linewidth}
{\small
\begin{tabular}{|l|r|r|r|r|r|r|}
\hline
\multirow{2}{*}{Dataset} & BFS & S$^*$ & TS$^*$ & BiBFS & HN & HL\\
\cline{2-7}
& \multicolumn{3}{c|}{$ms$} & \multicolumn{3}{c|}{$\mu s$}
\\\hline
Facebook & 1.9 & 0.5 & 19.6 & 61.2 & 45.7 & 19.9\\\hline
Slashdot & 1.7 & 0.7 & 46.8 & 31.4 & 20.3 & 1.5\\\hline
BerkStan & 0.3 & 2.1 & 206.7 & 36.1 & 10.6 & 3.8\\\hline
Youtube & 16 & 1.2 & 95 & 325.8 & 130.7 & 5.6\\\hline
As-skitter & 5.4 & 1.2 & 84.2 & 94.7 & 46.3 & 14\\\hline
Flickr & 45.2 & 2.9 & 182.1 & 3060 & 1825.2 & 79.1\\\hline
Flickr-growth & 71.9 & 3.7 & 332.5 & 1616.6 & 1219.6 & 103.6\\\hline
Wiki-talk & 21.7 & - & - & 58.3 & 14.2 & 1.1\\\hline
Orkut & 225.8 & 3.4 & 268 & 1372.9 & 1111.1 & 4639.5\\\hline
LiveJournal & 127.7 & - & - & 699.3 & 524 & -\\\hline
Twitter & 250.4 & - & - & 2384.3 & 2190.1 & 254.5\\\hline
\end{tabular}
}
\vspace{-2.0ex}
\caption{{\small Average Query Time on Random Query with Distance $\ge 4$}}
\label{tab:randlongpathquerytime}
\vspace*{-1.0ex}

\end{minipage}
\end{table*}

\begin{table*}[!ht]
\begin{minipage}[b]{0.42\linewidth}
{\small
\begin{tabular}{|l|r|r|r|r|r|}
\hline
\multirow{2}{*}{Dataset} & \multirow{2}{*}{BFS} & \multirow{2}{*}{BiBFS}  & \multirow{2}{*}{HN}  & \multicolumn{2}{c|}{HL} \\
\cline{5-6}
 & & & & \multicolumn{1}{l|}{HP-BBFS} & \multicolumn{1}{l|}{Join} \\ \hline

Facebook            &   30589     &   1723    &   1867   &   208   &  466  \\ \hline
Slashdot            &   41030 &   1380    &   1358   &   3     &  20  \\ \hline
BerkStan            &   11099     &   1462    &   405    &   78    &  39  \\ \hline
Youtube             &   505842    &   13941   &   6303   &   78    &  90  \\ \hline
As-skitter          &   161878    &   3580    &   1551   &   292   &  265  \\ \hline
Flickr              &   580315    &   36161   &   15494  &   1431  &  1330  \\ \hline
Flickr-growth       &   777994    &   23738   &   12412  &   2382  &  1431  \\ \hline
Wiki-Talk           &   1178526   &   4255    &   1111   &   1     &  7  \\ \hline
Orkut               &   1522640   &   29341   &   21954  &   71331 &  5367  \\ \hline
LiveJournal         &   1784211   &   14172   &   15554  &   -        &  -  \\ \hline
Twitter             &   3275797   &   55558   &   54884  &   13866 &  10757  \\ \hline

\end{tabular}
}
\vspace{-1.0ex}
\caption{{\small Average Search Space on Random Query}}
\label{tab:randomsearchspace}
\vspace*{-1.0ex}

\end{minipage}
\hspace{0.7cm}
\begin{minipage}[b]{0.48\linewidth}
{\small

\begin{tabular}{|l|r|r|r|r|r|r|r|}
    \hline
    \multirow{2}{*}{Dataset}&  \multicolumn{3}{c|}{Indexing Cost} & \multicolumn{3}{c|}{Preproc.Time(min)} \\
    \cline{2-7}
    & \multicolumn{1}{c|}{\tiny S$^*$(MB)} & \multicolumn{1}{c|}{\tiny HL$_{all}$(MB)}
    & \multicolumn{1}{c|}{{\tiny $\overline{|L(v)|}$}} &
    \multicolumn{1}{c|}{ S$^*$} & \multicolumn{1}{c|}{ HN} & \multicolumn{1}{c|}{ HL} \\ \hline

{ Facebook}	&	10	  & 955	&	8.2	&	3.2	&  2.2 &	3.8	\\	\hline
{ Slashdot}	&	26	  &	496	&	11.1	&	6.5	& 1.3  &	4.3	\\	\hline
{ BerkStan}	&	193	 &	291	&	21.6	&	64.3	& 0.3  &	1.7	\\	\hline
{ Youtube}	&	217	&  757	&	38.9	&	100.8	& 15.5 &	66	\\	\hline
{ As-skitter}	&	391	  &	1229	&	101.9	&	109.9	& 7.1	& 31.7	\\	\hline
{ Flickr}	&	626	&  	1536	&	232	&	163.8	& 43.4	& 202.5	\\	\hline
{ Flickr-growth}	  &	1004 &		4403.2	&	315.9	&	242.8	& 71.8 &	363.5\\\hline
{ WikiTalk}	&	-	&  	481	&	2.5	&	-	& 12.5	& 41.2	\\	\hline
{ Orkut}	&	8397	&  	13517	&	749.3	&	773.2	& 412.5	& 1431.6	\\	\hline
{ LiveJournal}	&	-	  &	-	&	-	&	-		& 334.2	& -\\\hline
{ Twitter} & - & 26931 & 464 & - & 233.9 & 390.2 \\ \hline

\end{tabular}
}
\vspace{-2.0ex}
\caption{{\small Preprocessing Cost on Random Query}}
\label{tab:precomp}
\vspace*{-1.0ex}

\end{minipage}
\end{table*}

\begin{table*}[!ht]
{\small
\begin{tabular}{|l|r|r|r|r|r|r|r|r|r|r|r|r|}
\hline
\multirow{2}{*}{Dataset} & \multicolumn{3}{c|}{$|H|=5000$} & \multicolumn{3}{c|}{$|H|=8000$} &
\multicolumn{3}{c|}{$|H|=10000$} & \multicolumn{3}{c|}{$|H|=15000$} \\
\cline{2-13}
&
$|H^\star|$ & $d_1(H)$ & $d_2(H)$ & $|H^\star|$ & $d_1(H)$ & $d_2(H)$ & $|H^\star|$ & $d_1(H)$ & $d_2(H)$ & $|H^\star|$ & $d_1(H)$ & $d_2(H)$
\\ \hline

Facebook      &   20854   &   247.7   &   217.1   &   27364   &   202.7   &   184.5   &   30554  &  182.2  &  168.1 &
36188 & 146.6 & 137.5\\
\hline
Slashdot      &   23359   &   204.5   &   179.5   &   27581   &   150.1   &   135.6   &   29500  &  128.4  &  117.2 &
32665 & 95.2 & 88.0\\
\hline
BerkStan      &   8290    &   769.3   &   177.8   &   16563   &   574.3   &   152.8   &   24618  &  492.8  &  138.1 &
34342 & 364.6 & 110.3\\
\hline
Youtube       &   49516   &   587.5   &   299.9   &   69474   &   429.9   &   254.9   &   76894  &  369.4  &  231.1 &
100595 & 279.2 & 189.3\\
\hline
As-skitter    &   41371   &   958.9   &   211.0   &   56245   &   701.3   &   184.8   &   64785  &  601.4  &  171.3 &
82439 & 453.0 & 146.3\\
\hline
Flickr        &   19198   &   2539.3  &   1433.3  &   32972   &   2005.8  &   1364.7  &   42312  &  1776.7 &  1295.0 &
63774 & 1403.7 & 1128.9\\
\hline
Flickr-growth &   22715   &   3175.3  &   1626.7  &   38819   &   2555.4  &   1615.5  &   49450  &  2284.0 &  1565.4 &
74569 & 1833.5 & 1407.7\\
\hline
Wiki-talk     &   24139   &   984.5   &   294.7   &   32435   &   669.2   &   220.3   &   36081  &  552.4  &  188.8 &
41567 & 385.9 & 139.9\\
\hline
Orkut         &   124607  &   3808.5  &   1720.9  &   189686  &   3022.9  &   1763.0  &   225678 &  2734.3 &  1763.4 &
319989 & 2305.0 & 1720.0\\
\hline
LiveJournal   &   151348  &   1172.3  &   702.1   &   229836  &   1004.5  &   673.4   &   278203 &  932.8  &  653.7 &
392423 & 808.8 & 611.0\\
\hline
Twitter       &   201521  &   9556.6  &   2877.8  &   346091  &   6762.9  &   2641.2  &   424853 &  5749.2 &  2463.3 &
564435 & 4267.5 & 2084.0\\
\hline

\end{tabular}
}
\vspace{-2.0ex}
\caption{{\small Hub-Network Statistics}}
\label{tab:subgraphstat}
\vspace*{-1.0ex}
\end{table*}


The benchmarking datasets are listed in Table~\ref{realdata}.
Most of them are gathered from online social networks, with the number of vertices ranging from several tens of thousands to more than  $10$ million.
Others also exhibit certain properties commonly observed in social networks, such as small diameter and relatively high average vertex degree.
All datasets are downloadable from Stanford Large Network Dataset Collection~\footnote{http://snap.stanford.edu/data/index.html}, Max Planck Institute's Online Social Network Research Center~\footnote{http://socialnetworks.mpi-sws.org/}, and Social Computing Data Repository at Arizona State University~\footnote{http://socialcomputing.asu.edu/datasets/}. 

In Table~\ref{realdata}, we present important characteristics of all real datasets,
where $\overline{\delta}$ is average vertex degree (i.e., $2|E|/|V|$) and $d_{0.9}$ is 90-percentile effective diameter~\cite{Leskovec:2005:GOT:1081870.1081893}.
Finally, in the experimental study, we focus on the $6$-degree shortest path queries ($k=6$) as they are the most commonly used and also the most challenging one.

\comment{
Here, we describe more details of datasets in the following.


\noindent {\bf Facebook:} a social network crawled from Facebook. Each node represents a user and there is a directed edge from node $i$ to node $j$ if user $i$ posted a message on user $j$'s personal page.

\noindent {\bf Slashdot:} a technology-related social network. It contains user-to-user friendship links which were collected in November 2008.

\noindent {\bf BerkStan:} a web graph describing the hyperlinks from pages of berkely.edu to pages of stanford.edu.
These data were collected in 2002 and presented in~\cite{DBLP:journals/corr/abs-0810-1355}.


\noindent {\bf Youtube:} a YouTube video-sharing network containing around $1.1$ million users and $5$ million user-to-user links.

\noindent {\bf As-skitter:} an internet topology graph from traceroutes run daily in 2005.

\noindent {\bf Flickr:} a Flickr photo-sharing network with around $1.8$ million users and $22$ million links, which were crawled in 2007.

\noindent {\bf Flickr-growth:} another Flickr social network used to investigate the behavior of social network growth in~\cite{mislove-2008-flickr}.

\noindent {\bf WikiTalk:} a Wikipedia users' communication network. This network contains all the users and discussion from the inception of Wikipedia till January 2008.
In the network, each node represents one user and there is a directed edge from node $i$ to node $j$ if user $i$ edited a talk page of user $j$ for communication.

\noindent {\bf Orkut:} an online social network owned by Google. Each node denotes one user and each undirected link identifies friendship between two users.


\noindent {\bf LiveJournal:} a fraction of the LiveJournal social network. Each node is a highly active user and undirected edges are used to identify the friendship between different users.



}


%

\vspace*{-2.0ex}
\subsection{Experimental Results}
\label{results}

In the following, we report effectiveness and efficiency of the shortest path computation algorithms from different perspectives:  

\noindent{\bf Query Results on Random Queries}
In this experiment, we randomly generate $10,000$ vertex pairs with various distances and execute all algorithms on these queries to study their performance. Here, we select $10,000$ vertices with highest vertex degree as hubs. 
Table~\ref{tab:randquerytime} presents the average query time for $10,000$ queries on all the methods and Table~\ref{tab:randlongpathquerytime} highlights the average query time for those vertex pairs whose distance is no less than $4$ (longer path) as these are the more challenging ones (the longer the path, the likely more hubs will be expanded).
Note that for BFS and two sketch methods Sketch(S$^\ast$) and TreeSketch(TS$^\ast$), we use the {\em millisecond} ($10^{-3}$) as the unit, as they typically have much longer query time, and for BiBFS and our new approaches, Hub-Network (HN) and Hub$^2$-Labeling (HL) approaches, we use the {\em microsecond} ($10^{-6}$) as the unit, as they are much faster. 
Their corresponding average search space per query is reported in Table~\ref{tab:randomsearchspace},
where column ``HP-BBFS'' records the average number of vertices visited by HP-BBFS (Hub-Pruning Bidirectional BFS) in Hub$^2$-Labeling (HL) and column ``Join'' records the average times of pairwise join on the core-hubs labeling $L(s)$ and $L(t)$ in HL.
We make the following observations on the query time and average search space:

\noindent 1) The Hub$^2$-Labeling (HL) is clearly the winner among all algorithms, which is on average more than $2000$ times faster than BFS. In most of the social networks, like As-skitter and WikiTalk, the average query time of Hub$^2$-Labeling (HL) is only tens of microseconds ($10^{-8}$ second), and except for one (Orkut), all of tham are less than $1$ms.  Overall, Hub$^2$-Labeling (HL) is on average $23$ times faster than BiBFS.
Specifically, we observe  that compared to BiBFS, the Hub-Pruning Bidirectional Search (HP-BBFS) of  achieves significant improvement in terms of search space, which is around $800$ times smaller than BiBFS (Table~\ref{tab:randomsearchspace}).

\noindent 2) The Hub-Network (HN) is on average about $2$ times faster than BiBFS (with no additional storage cost but reorganizes the network structure).  It is about two orders of magnitude faster than BFS but is about $10$ times slower than the Hub$^2$-Labeling approach. 
  
\noindent 3) Sketch (S$^\star$) is on average about $10$ times faster than BFS but it fails to run on a few datasets. 
The TreeSketch (TS$^\star$) is on average $70$ times slower than Sketch. Both Hub-Network and Hub-Labeling approaches are are on average more than two orders of magnitude faster than Sketch, the fastest approximation method.

\noindent 4) For long distance queries $d(u,v) \geq 4$ the exact shortest path approaches require longer query time (Table~\ref{tab:randlongpathquerytime}). However, the increase for the Hub-Network (HN) and Hub$^2$-Labeling (HL) are smaller than BFS and BiBFS. Also, it is interesting to observe the approximate shortest path approaches do not show performance decrease though both of them are still very slow.

\noindent{\bf Preprocessing Cost:}
Table~\ref{tab:precomp} shows preprocessing cost of the Sketch-based approach along with HL, consisting of indexing size and precomputation time. 
The first column S$^\star$ records the index size  (MB) for the Sketch method.
The second column HL$_{total}$ records total index size of Hub$^2$-Labeling (HL), which is the sum of core-hubs labeling cost and distance matrix size.
Column $\overline{|L(v)|}$ record the average number of core-hubs stored by each vertex. 
Remarkably, the core-hub labeling scheme in Hub$^2$-Labeling (HL) is very effective, as there is a very small portion of core-hubs recorded by each vertex.
In most of the network, the average number of core-hubs per vertex is no more than $2\%$ of the total hubs. 
In particular, for network WikiTalk, only $2.5$ core-hubs are stored in each vertex on average, which potentially leads to efficient query answering.
However, for LiveJournal, the Hub$^2$-Labeling is too expensive to be materialized in the main memory. 
In terms of precomputation time, Hub$^2$-Labeling can be constructed faster than Sketch on $7$ out of $10$ networks. The construction time of HubNetwork (HN) is average more than three times faster than the Hub$^2$-Labeling (HL), and it does not need any additional memory cost. 

\comment{
In this experiment, we evaluate the performance of HL against BiBFS, BFS and Sketch-based approximation approaches on queries with specific distance.
For each network, we randomly generate $6$ sets of queries, such that vertex pairs in each query set has specific distance, ranging from $2$ to $6$, and $7$ indicates the group of queries with the vertex pair distances are greater than $6$.
We report the average query time of BFS, BiBFS, HL, Sketch and TreeSketch on each query set in Table~\ref{fig:distquerytime}.
Here we only report Sketch and TreeSketch in the Sketch-based methods as the former has the fastest query time and the later has the best approximation accurracy.
We still select $10,000$ vertices with largest degree to be hub-set $\mathcal{H}$.

In all these benchmarking networks, we observe that HL clearly outperforms all other approaches  for queries with their distances no less than $3$.
Online search algorithms BFS and BiBFS turn out to be effective on short distance queries as the search space is generally limited and there is not much need for the distance estimation in the Step $1$ for HL.
For queries with relatively large distance (i.e., $\ge 5$), HL provides a nice speedup over BiBFS and BFS.
In several networks, the speedup over BiBFS can be close to two orders of magnitude.
This demonstrates the effectiveness of HBBFS by avoiding expanding the hubs in the network.


The performances of BFS and BiBFS  exhibit the climbing trend as the distances on the query pairs increase.
As the query distance increases, the performance of BFS drops rather quickly and become much slower than both Sketch and TreeSketch on queries with distance greater than or equal to $6$.
In the meantime, the performance of HL and the two approximation approaches Sketch and TreeSketch are quite stable with respect to the query distance.
Overall, HC is faster than Sketch and TreeSketch  on average by two orders of magnitude and three orders of magnitude, respectively.
Furthermore, similar to the results on random queries (with various distances), BiBFS achieves much better query performance than Sketch on most of networks.

\vspace*{-2.0ex}}

\noindent{\bf Impacts of Hub Number:}
In this experiment, we study the effect of different number of hubs on query performance. 
Here, we vary the hub-set size from $5,000$ to $15,000$ and conduct the experiment on $10,000$ randomly generated queries with various distances.
Table~\ref{tab:hubsize} shows the average query time of Hub-Network (HN) and Hub$^2$-Labeling approaches using different number of hubs. 
In most of these networks, the best query performance is achieved when the number of hubs lies between $10K$ and $15K$.
Though a large number of hubs may potentially help reduce the search space of the bidirectional search in Hub$^2$-Labeling (HL), it may also increase the size of core-hubs associated with each vertex.
We observe that the query performance obtained by using $10K$ hub is comparable to the best one). 
Note that here due to space limitation, we do not report the detailed precomputation cost in terms of construction time and index size (for Hub$^2$-Labeling). Overall, as the number of hub increases, most large networks, show an increasing trend regarding the average index size. Interestingly, when hub-set size increases, significant reduction of average index size is observed on WikiTalk. This is in part explained by its very small diameter. In terms of the precomputation time, as more hubs are chosen, the computational cost of Hub-Network and Hub$^2$-Labeling becomes larger, because more BFS needs to performed. Indeed, the precomputation time increases almost linearly with respect to the hub-set size.

\noindent{\bf Hub-Network Statistics:} 
Finally, we report the basic statistics of the discovered distance preserving Hub-Network. Specifically, we are introduced in two following two questions: 1) given a set of hubs, how large the hub-network will be? What is the size of $|H^\star|$?
2) what are the degree difference between the hubs in the original network and in the Hub-Network? Do we observe a significant degree decreasing?
To answer these two questions,  in Table~\ref{tab:subgraphstat}, we report $|H^\star|$ (the number of total vertices in the hub-network), $d_1(H)$ the average degree of hubs in the original graph, and $d_2(H)$, the average degree of hubs in the extracted hub-network, with respect to $5K$, $8K$, $10K$ and $15K$ hubs. 
We observe for most graphs, the size of $|H^\star|$ is a few times larger than the hub number; however, for Orkut, LiveJournal, and Twitter, the hub network becomes quite large at $10K$ and $15K$ hubs. 
Also, in general, the degree of hubs in the hub-network has been lowered and on several graphs, the average degree is reduced smaller than 1/3 of the original average degree. We also observe that the ability of lowering degree is correlated with the search performance: the better the hub degree is lowered, the better query performance improvement we can get from the Hub-Network based bidirectional BFS.

\comment{
\subsection{Query Efficiency}
\label{queryeffect}

In this section, we study the query performance of our core-based search algorithm, bidirectional breadth-first search and all Sketch-based algorithms.
First, we assess the efficiency of query procedures on two different query sets: completely random shortest path queries and query set with specific distance.
Then, we investigate the effect of different core sizes on query performance.

[To help understand the benefit of using core-based algorithm, we consider record the search space reduction compared to BiBFS]

\subsubsection{Random queries}
\label{randomquery}

\subsubsection{Queries with specific distance}
\label{distquery}

\subsubsection{The effect of core size}
\label{coresizeeffect}

In this group of experiment, we study the effect of core size on the performance of our query algorithm.
Here, we vary the core size from $10,000$ to $20,000$ on the real graphs.

[10K 12K 14K 16K 18K 20K]

\subsection{Preprocessing}
\label{preproeffect}

core-based algorithm:
[number of routing entries]
[size of core path tables]

sketch algorithm:
[index size]
}

\comment{
\subsection{Approximation Quality}
\label{approxeffect}

In this section, we study our approximate version of core-based algorithm (i.e., {\bf CoreApprox}) in terms of query time and approximation Error.
For each query $(s,t)$, their approximation error is expressed as relative error ratio between the result produced by algorithms $l(s,t)$ and the true shortest distance $dist(s,t)$:

\[ e(s,t) = \frac{|l(s,t)-dist(s,t)|}{dist(s,t)}  \]

We calculate the average approximation error over every query in one query set.

[query time and approximation error]

[effect of core size]
}

\vspace*{-2.0ex}
\section{Conclusion}
\label{conc}

In this paper, we introduce a set of novel techniques centered on hubs for $k$-degree shortest path computation in large social networks. The Hub-Network and Hub$^2$-Labeling algorithms can help significantly reduce the search space. 
The extensive experimental study demonstrates that these approaches can handle very large networks with millions of vertices, and its query processing is much faster than online searching algorithms and Sketch-based approaches, the state-of-the-art shortest path approximation algorithms.
To the best of our knowledge, this is the first practical study on computing exact shortest paths on large social networks.
In the future, we will  study how to parallelize the index construction and query answering process. We also plan to investigate how to compute $k$-degree shortest path on dynamic networks.


\begin{small}
\bibliographystyle{plain}
\bibliography{./bib/cikm,./bib/reachability,./bib/3hop,./bib/reachpaper,./bib/labelreachability,./bib/Yangdissertation,./bib/distance10,./bib/ComplexNetwork,./bib/simplification,./bib/ComplexNetwork2,./bib/socialnetwork,./bib/distance,./bib/graphdb}
\end{small}

\end{document}